\documentclass{article}
\usepackage{color}
\def\vol#1#2#3{{\bf {#1}} ({#2}) {#3}}
\def\NP{Nucl.~Phys. }
\def\PL{Phys.~Lett. }
\def\PR{Phys.~Rev. }
\def\PRP{Phys.~Rep. }
\def\PRL{Phys.~Rev.~Lett. }
\def\PTP{Prog.~Theor.~Phys. }

\def\IJMP{Int.~J.~Mod.~Phys. }

\def\no{\nonumber}

\def\2tvec#1#2{
\left(
\begin{array}{c}
#1  \\
#2  \\   
\end{array}
\right)}

\def\mat2#1#2#3#4{
\left(
\begin{array}{cc}
#1 & #2 \\
#3 & #4 \\
\end{array}
\right)
}

\def\Mat3#1#2#3#4#5#6#7#8#9{
\left(
\begin{array}{ccc}
#1 & #2 & #3 \\
#4 & #5 & #6 \\
#7 & #8 & #9 \\
\end{array}
\right)
}

\def\3tvec#1#2#3{
\left(
\begin{array}{c}
#1  \\
#2  \\   
#3  \\
\end{array}
\right)}

\def\4tvec#1#2#3#4{
\left(
\begin{array}{c}
#1  \\
#2  \\   
#3  \\
#4  \\
\end{array}
\right)}

\def\L{\left}
\def\R{\right}

\def\pl{\partial}

\def\hbar{\hspace{1mm}\bar{}\hspace{-1mm}h}

\def\eqn#1{
\begin{eqnarray}
#1
\end{eqnarray}
}

\begin{document}
  \title{\bf Phenomenology of $S_4$ Flavor Symmetric extra U(1) model}

  \author{Yasuhiro Daikoku\footnote{E-mail: yasu\_daikoku@yahoo.co.jp}, 
\quad Hiroshi Okada\footnote{E-mail: hokada@kias.re.kr}
\\
  {\em Institute for Theoretical Physics, Kanazawa University, Kanazawa
  920-1192, Japan.}$^*$$^\ddagger$ \\
  {\em School of Physics, KIAS, Seoul 130-722, Korea}$^\dagger$}
\maketitle

\begin{abstract}
We study several phenomenologies of an $E_6$ inspired extra U(1) model with $S_4$ flavor symmetry.
With the assignment of left-handed quarks and leptons to $S_4$-doublet, SUSY flavor problem
is softened. As the extra Higgs bosons are neutrinophilic, 
baryon number asymmetry in the universe is
realized by leptogenesis without causing gravitino overproduction.
We find that the allowed region for the lightest chargino mass is given by 100-140 GeV,
if the dark matter is a singlino dominated neutralino whose mass is about 36 GeV.

\end{abstract}

\vspace{-11.5cm}
\begin{flushright}
KIAS-P13019
\end{flushright}
\newpage



\section{Introduction}

Standard model (SM) is a successful theory of gauge interactions, 
however there are many unsolved puzzles in  the Yukawa sectors.
What do the Yukawa hierarchies of quarks and charged leptons mean?
Why is the neutrino mass so small? Why does the generation exist?
These questions give  rise to the serious motivation to extend SM. 
Another important puzzle of SM is the existence of
large hierarchy between electroweak scale $M_W\sim 10^2$GeV and Planck scale
$M_P\sim 10^{18}$GeV. The elegant solution of this hierarchy problem 
is supersymmetry (SUSY)\cite{SUSY}.
Recent discovery of the Higgs boson at the Large Hadron Collider (LHC)
may suggest the existence of SUSY because the mass of Higgs boson;
$125-126$ GeV \cite{higgs}, is in good agreement with the SUSY prediction.
Moreover, in the supersymmetric model, more information 
are provided for the Yukawa sectors.

In the supersymmetric model, the Yukawa interactions are introduced
in the form of superpotential. Therefore, to understand the structure of the Yukawa interaction,
we have to understand the structure of superpotential. 
In the minimal supersymmetric standard model (MSSM),
as the Higgs superfields $H^U$ and $H^D$ are vector-like under the SM gauge symmetry
$G_{SM}=SU(3)\times SU(2)\times U(1)$, we can introduce $\mu$-term;
\eqn{
\mu H^U H^D,
}
in superpotential. The natural size of parameter $\mu$ is $O(M_P)$, however
$\mu$ must be $O(M_W)$ to succeed in breaking electroweak gauge symmetry.
This is so-called $\mu$-problem.
The elegant solution of $\mu$-problem is to make Higgs superfields chiral
under a new $U(1)_X$ gauge symmetry. Such a model is achieved based on 
$E_6$-inspired extra U(1) model \cite{extra-u1}.  
The new gauge symmetry replaces the $\mu$-term by trilinear term;
\eqn{
\lambda SH^UH^D,
}
which is converted into effective $\mu$-term when singlet $S$ develops $O(1\mbox{TeV})$ 
vacuum expectation value (VEV) \cite{mu-problem}.
At the same time, the baryon and lepton number violating terms in MSSM are replaced by
single G-interactions;
\eqn{
GQQ+G^cU^cD^c+GU^cE^c+G^cQL,
}
where $G$ and $G^c$ are new colored superfields which must be introduced to cancel gauge anomaly.
These terms induce very fast proton decay. To make proton stable, we must tune
these trilinear coupling constants to be very small $\sim O(10^{-14})$, which gives rise to  a new puzzle. 

The existence of small parameters in superpotential suggests that a new symmetry is hidden.
As such a symmetry suppresses the Yukawa coupling of 
the first and the second generation of the quarks and the charged leptons,
it should be flavor symmetry.
We guess several properties that the flavor symmetry should have {in order.
At first, the flavor symmetry should be non-abelian and include triplet representations,
which is the simple reason why three generations exist.
At second, remembering that the quark and the charged-lepton masses are
suppressed by $SU(2)_W$ gauge symmetry as the left-handed fermions 
are assigned to be doublet and the right-handed fermions are assigned to be singlet,
the flavor symmetry should include doublets.
In this case, if we assign the first and the second generation of the left-handed quarks and leptons
to be doublets and the right-handed to be singlets, then suppression of Yukawa couplings
is realized in the same manner as $SU(2)_W$. At the same time, this assignment
softens the SUSY-flavor problem because of the left-handed sfermion mass degeneracy.
Finally, any products of the doublets should not include the triplets.
In this case, we can forbid single G-interactions when we assign $G$ and $G^c$ to be triplets and 
the others to be doublets or singlets. 
The existence of triplets $G$ and $G^c$ compels
all fermions to consist of three generations to cancel gauge anomaly.
As one of the candidates of the flavor symmetries which have the nature
as above, we consider $S_4$ \cite{s4}.
In such a model,  the generation structure is understood as a new system to
stabilize proton \cite{s4u1}.

In section 2, we introduce new symmetries and explain how to break them.
In section 3, we discuss Higgs multiplets. In section 4, we give order-of-magnitude estimates of 
the mass matrices of quarks and leptons and flavor changing processes.
In section 5, we discuss cosmological aspects of our model.
Finally, we give conclusions in section 6.



\section{Symmetry Breaking}

At first we introduce new symmetries and explain how to break
these symmetries. The charge assignments of the superfields
are also defined in this section.

\subsection{Gauge symmetry}

We extend the gauge symmetry from $G_{SM}$ to $G_{32111}=G_{SM}\times U(1)_X\times U(1)_Z$,
and add new superfields $N^c,S,G,G^c$ which are embedded in {\bf 27} representation of
$E_6$ with quark, lepton superfields $Q,U^c,D^c,L,E^c$ and Higgs superfields $H^U,H^D$.
Where $N^c$ is right-handed neutrino (RHN), $S$ is $G_{SM}$ singlet and $G,G^c$ are
colored Higgs. The two U(1)s are linear combinations of 
$U(1)_\psi,U(1)_\chi$ where 
$E_6\supset SO(10)\times U(1)_\psi\supset SU(5)\times U(1)_\chi\times U(1)_\psi$,
and their charges $X$ and $Z$ are given as follows
\eqn{
X=\frac{\sqrt{15}}{4}Q_\psi+\frac14 Q_\chi,\quad
Z=-\frac14 Q_\psi+\frac{\sqrt{15}}{4}Q_\chi. 
}
The charge assignments of the superfields are given in Table 1.
To break $U(1)_Z$, we add new vector-like superfields $\Phi,\Phi^c$ where
$\Phi^c$ is the same representation as RHN under the $G_{32111}$ and its anti-representation
$\Phi$ is originated in ${\bf 27}^*$.
To discriminate between $N^c$ and $\Phi^c$, we introduce $Z^R_2$ symmetry and
assign $\Phi^c,\Phi$ to be odd.
The invariant superpotential under these symmetries is given by
\eqn{
W_{32111}&=&W_0+W_S+W_G+W_\Phi , \\
W_0&=&Y^UH^UQU^c+Y^DH^DQD^c+Y^LH^DLE^c+Y^N H^ULN^c+\frac{Y^M}{M_P}\Phi\Phi N^cN^c , \\
W_S&=&kSGG^c+\lambda SH^UH^D , \\
W_G&=&Y^{QQ} GQQ+Y^{UD} G^cU^cD^c+Y^{UE} GU^cE^c+Y^{QL} G^cQL+Y^{DN}GD^cN^c , \\
W_\Phi&=&M_\Phi \Phi\Phi^c+\frac{1}{M_P}Y^\Phi(\Phi\Phi^c)^2,
}
where unimportant higher dimensional terms are omitted\footnote{As the 4-th and 5-th order terms of 27 representation of $E_6$
are forbidden  by gauge symmetry, 
the leading order terms of higher dimensional operators are 6-th order which are harmless to
proton stability. The single $\Phi$ dressed terms such as $\Phi G^cU^cU^cE^c$ are forbidden by
$Z^R_2$ symmetry.}. Since the interactions $W_S$ 
drive squared mass of $S$ to be negative through renormalization group equations (RGEs),
spontaneous $U(1)_X$ symmetry breaking is realized and
$U(1)_X$ gauge boson $Z'$ acquires the mass
\eqn{
m(Z')=5\sqrt{2}g_x\L<S\R>=5\sqrt{2}\L(\frac{1}{2\sqrt{6}}g_X\R)\L<S\R>=0.5255\L<S\R>,
}
where the used value $g_X(M_S=1\mbox{TeV})=0.3641$
is calculated based on the RGEs given in appendix A,
and $\L<H^{U,D}\R>\ll \L<S\R>$ is assumed based on the experimental constraint \cite{zprime2011}
\eqn{
m(Z')>1.52\mbox{TeV} ,
}
which imposes lower bound on VEV of $S$ as
\eqn{
\L<S\R> > 2892 \mbox{GeV}.
}

To drive squared mass of $\Phi^c$ to be negative, we introduce 4th generation
superfields $H^U_4,L_4$ and their anti-representations $\bar{H}^U_4,\bar{L}_4$
and add new interaction
\eqn{
W\supset Y^{LH}\Phi^cH^U_4L_4.
} 
To forbid the mixing between 4th generation and three generations,
we introduce 4-th generation parity $Z^{(4)}_2$ and
assign all 4-th generation superfields to be odd.
If $M_\Phi=0$ in $W_\Phi$, then $\Phi,\Phi^c$ develop large VEVs 
along the D-flat direction of $\L<\Phi\R>=\L<\Phi^c\R>=V$, 
$U(1)_Z$ is broken and  $U(1)_Z$ gauge boson $Z''$ acquires the mass
\eqn{
m(Z'')=8g_zV=8\L(\frac16\sqrt{\frac52}g_Z\R)V=0.9202V,
}
where the used value $g_Z(\mu=M_I)=0.4365$ is calculated by the same way as $g_X$.
We determine the values of two gauge couplings $g_X,g_Z$ by requiring 
three $U(1)$ gauge coupling constants are unified at reduced Planck scale 
$M_P=2.4\times 10^{18}\mbox{GeV}$ as
\eqn{
g_Y(M_P)=g_X(M_P)=g_Z(M_P).
}
In this paper we fix the VEV as
\eqn{
V=M_I=10^{11.5}\mbox{GeV}.
}
 RHN obtains the mass
\eqn{
M_R\sim \frac{V^2}{M_P}\sim 10^{4-5} \mbox{GeV},
}
through the quartic term in $W_0$. 

After the gauge symmetry breaking, since the R-parity symmetry defined by
\eqn{
R=Z^R_2\exp\L[\frac{i\pi}{20}(3x-8y+15z)\R] ,
}
remains unbroken, the lightest SUSY particle (LSP) is a promising candidate for cold dark matter.
As we adopt the naming rule of superfields
as the name of superfield is given by its R-parity even component, we call $G, G^c$ "colored Higgs".

Before considering flavor symmetry, we should keep in mind following points.
As the interaction $W_G$ induces too fast proton decay, they must be strongly suppressed. 
As the mass term $M_\Phi\Phi\Phi^c$ prevents $\Phi,\Phi^c$ from developing VEV and breaking
$U(1)_Z$ symmetry, it must be forbidden.
In $W_0$, the contributions to flavor changing processes from the extra Higgs bosons must be
suppressed \cite{u1fv}.

\begin{table}[htbp]
\begin{center}
\begin{tabular}{|c|c|c|c|c|c|c|c|c|c|c|c||c|c|}
\hline
                   &$Q$ &$U^c$    &$E^c$&$D^c$    &$L$ &$N^c$&$H^D$&$G^c$    &$H^U$&$G$ &$S$ &$\Phi$&$\Phi^c$\\ \hline
$SU(3)_c$          &$3$ &$\bar{3}$&$1$  &$\bar{3}$&$1$ &$1$  &$1$  &$\bar{3}$&$1$  &$3$ &$1$ &$1$   &$1$     \\ \hline
$SU(2)_w$          &$2$ &$1$      &$1$  &$1$      &$2$ &$1$  &$2$  &$1$      &$2$  &$1$ &$1$ &$1$   &$1$     \\ \hline
$y=6Y$             &$1$ &$-4$     &$6$  &$2$      &$-3$&$0$  &$-3$ &$2$      &$3$  &$-2$&$0$ &$0$   &$0$     \\ \hline  
$6\sqrt{2/5}Q_\psi$&$1$ &$1$      &$1$  &$1$      &$1$ &$1$  &$-2$ &$-2$     &$-2$ &$-2$&$4$ &$-1$  &$1$     \\ \hline
$2\sqrt{6}Q_\chi$  &$-1$&$-1$     &$-1$ &$3$      &$3$ &$-5$ &$-2$ &$-2$     &$2$  &$2$ &$0$ &$5$  &$-5$   \\ \hline 
$x=2\sqrt{6}X$     &$ 1$&$ 1$     &$ 1$ &$2$      &$2$ &$0 $ &$-3$ &$-3$     &$-2$ &$-2$&$5$ &$0$   &$0$     \\ \hline 
$z=6\sqrt{2/5}Z$   &$-1$&$-1$     &$-1$ &$2$      &$2$ &$-4$ &$-1$ &$-1$     &$ 2$ &$ 2$&$-1$&$4$   &$-4$   \\ \hline
$Z^R_2$            &$+$ &$+$      &$+$  &$+$      &$+$ &$+$  &$+$  &$+$      &$+$  &$+$ &$+$ &$-$   &$-$     \\ \hline 
$R$                &$-$ &$-$      &$-$  &$-$      &$-$ &$-$  &$+$  &$+$      &$+$  &$+$ &$+$ &$+$   &$+$     \\ \hline
\end{tabular}
\end{center}
\caption{$G_{32111}$ assignment of superfields. Where the $x$, $y$ and $z$ are charges of
$U(1)_X$, $U(1)_Y$ and $U(1)_Z$, and $y$ is hypercharge. The charges of $U(1)_\psi$ and
$U(1)_\chi$ which are defined in Eq.(4) are also given.}
\end{table}

\subsection{$S_4$ flavor symmetry}

If we introduce $S_4$ flavor symmetry and
assign $G,G^c$ to be triplets, then $W_G$ defined in Eq.(8) is forbidden. This is
because any products of doublets and singlets of $S_4$ do not contain triplets.
The multiplication rules of representations of $S_4$ are given in appendix B.
Note that we assume full $E_6$ symmetry does not realize at Planck scale,
therefore there is no need to assign all superfields to the same flavor representations.
In this model the generation number three is imprinted in $G, G^c$.
Therefore they may be called "G-Higgs" (generation number imprinted colored Higgs).

Since the existence of G-Higgs which has life time longer than 0.1 second spoils the
success of Big Ban nucleosynthesis (BBN)\cite{reheating}, $S_4$ symmetry must be broken.
Therefore we assign $\Phi$ to be triplet and $\Phi^c$ to be doublet and singlet to forbid
$M_\Phi\Phi\Phi^c$.
With this assignment, $S_4$ symmetry is broken due to the VEV of $\Phi$ and the
effective trilinear terms are induced by pentatic terms
\eqn{
W_{NRG}=\frac{1}{M^2_P}\Phi\Phi^c\L(GQQ+G^cU^cD^c+GU^cE^c+G^cQL+GD^cN^c\R).
}
The size of effective coupling constants of these terms is given by
\eqn{
\frac{\L<\Phi\R>\L<\Phi^c\R>}{M^2_P}\sim \frac{M_R}{M_P} \sim 10^{-14}.
}
This is the marginal size to satisfy the BBN constraint \cite{f-extra-u1}.
This relation gives the information about the RHN mass scale if the life time of G-Higgs is measured.

The assignments of the other superfields are determined based on following criterion,
(1)The quark and charged lepton mass matrices reproduce observed mass hierarchies and
CKM and MNS matrices. (2)The third generation Higgs $H^U_3,H^D_3$ are specified as MSSM Higgs
and the first and second generation Higgs superfields $H^U_{1,2}$ are neutrinophilic
which are needed for successful leptogenesis. 

To realize Yukawa hierarchies, we introduce gauge singlet and $S_4$ doublet flavon superfield $D_i$
and fix the VEV of $D_i$ by
\eqn{
V_D=\sqrt{|\L<D_1\R>|^2+|\L<D_2\R>|^2}=0.1M_P=2.4\times 10^{17} \mbox{GeV},
}
then the Yukawa coupling constants are expressed in the power of the parameter
\eqn{
\epsilon=\frac{V_D}{M_P}=0.1,
}
which is realized by $Z_{17}$ symmetry\footnote{For $Z_{N+3}$ symmetry ($N>0$),  the scale of $\epsilon$ is given by
$\epsilon\sim (m_{SUSY}/M_P)^{1/(N+1)} \sim 10^{-15/(N+1)}$.
To give $\epsilon=0.1$, we have to choose $N\sim 14$. In the case of $N=13$,
the superpotential has dangerous F-flat direction as $D^2_1+D^2_2=0$.
Therefore we have to select $N=14$ or $N=15$. In this paper, we selsect $N=14$ and $Z_{17}$ symmetry.}.
To drive the squared mass of flavon to be negative, we add 5-th and 6-th generation
superfields $L_{5,6},D^c_{5,6}$ as $S_4$-doublets and their anti-representations
$\bar{L}_{5,6},\bar{D}^c_{5,6}$ and introduce trilinear terms as
\eqn{
W_5&=& Y^{DD}[D_1(D^c_5\bar{D}^c_6+D^c_6\bar{D}^c_5)
+D_2(D^c_5\bar{D}^c_5-D^c_6\bar{D}^c_6)]\no \\
&+&Y^{LL}[D_1(L_5\bar{L}_6+L_6\bar{L}_5)+D_2(L_5\bar{L}_5-L_6\bar{L}_6)],
}
where the mass scale of these fields is given by
\eqn{
M_{L_5}=Y^{DD}V_D=Y^{LL}V_D
=\epsilon M_P=2.4\times 10^{17}\mbox{GeV}.
}
We assign the 5th and 6-th generation superfields to be $Z^{(5)}_2$-odd.
The representation of all superfields under the flavor symmetry is given in Table 2.
The mass terms of 4-th generation fields are given by
\eqn{
W_4=Y^{LH}\Phi^c_3L_4H^U_4+Y^L\frac{(D^2_1+D^2_2)^2}{M^3_P}L_4\bar{L}_4
+Y^H\frac{(D^2_1+D^2_2)^2}{M^3_P}H^U_4\bar{H}^U_4,
}
where
\eqn{
M_{L_4}=\epsilon^4Y^LM_P=\epsilon^4Y^HM_P=2.2\times 10^{14}\mbox{GeV},
}
which realizes gauge coupling unification at Planck scale as
\eqn{
g_3(M_P)=g_2(M_P).
}

\begin{table}[htbp]
\begin{center}
\begin{tabular}{|c|c|c|c|c|c|c|c|c|c|}
\hline
         &$Q_i$     &$Q_3$      &$U^c_1$      &$U^c_2$   &$U^c_3$   &$D^c_1$   &$D^c_2$    &$D^c_3$   
         &$L_i$     \\
      \hline
$S_4$    &${\bf 2}$ &${\bf 1}$  & ${\bf 1}$   &${\bf 1}$ &${\bf 1}$ &${\bf 1}$ &${\bf 1'}$ &${\bf 1}$ 
         &${\bf 2}$ \\
      \hline
$Z^{(2)}_2$&$+$     &$+$        &$+$          &$+$       &$+$       &$+$       &$+$        &$+$ 
         &$+$       \\  
      \hline
$Z^N_2$  &$+$       &$+$        &$+$          &$+$       &$+$       &$+$       &$+$        &$+$ 
         &$+$       \\  
      \hline      
$Z_{17}$ &$2/17$    &$0$        &$4/17$       &$1/17$    &$0$       &$3/17$    &$2/17$     &$2/17$
         &$2/17$    \\ 
      \hline
$Z^R_2$  &$+$       &$+$        &$+$          &$+$       &$+$       &$+$       &$+$        &$+$ 
         &$+$       \\  
      \hline
$Z^{(4)}_2$&$+$     &$+$        &$+$          &$+$       &$+$       &$+$       &$+$        &$+$
         &$+$       \\  
      \hline
$Z^{(5)}_2$&$+$     &$+$        &$+$          &$+$       &$+$       &$+$       &$+$        &$+$
         &$+$       \\  
      \hline      
      \hline
         &$L_3$     &$E^c_1$    &$E^c_2$      &$E^c_3$   &$N^c_1$   &$N^c_2$   &$N^c_3$    &$H^U_i$
         &$H^U_3$   \\
      \hline
$S_4$    &${\bf 1}$ &${\bf 1}$  &${\bf 1}$    &${\bf 1}$ &${\bf 1}$ &${\bf 1}$ &${\bf 1}$  &${\bf 2}$
         &${\bf 1}$ \\
      \hline
$Z^{(2)}_2$&$+$     &$+$        &$+$          &$+$       &$-$       &$-$       &$-$        &$-$
         &$+$       \\  
      \hline 
$Z^N_2$  &$+$       &$+$        &$+$          &$+$       &$-$       &$+$       &$+$        &$+$ 
         &$+$       \\  
      \hline      
$Z_{17}$ &$2/17$    &$3/17$     &$1/17$       &$0$       &$0$       &$0$       &$0$        &$1/17$ 
         &$0$       \\ 
      \hline
$Z^R_2$  &$+$       &$+$        &$+$          &$+$       &$+$       &$+$       &$+$        &$+$ 
         &$+$       \\  
      \hline
$Z^{(4)}_2$&$+$     &$+$        &$+$          &$+$       &$+$       &$+$       &$+$        &$+$
         &$+$       \\  
      \hline
$Z^{(5)}_2$&$+$     &$+$        &$+$          &$+$       &$+$       &$+$       &$+$        &$+$
         &$+$       \\  
      \hline      
      \hline
         &$H^D_i$   &$H^D_3$    &$S_i$        &$S_3$     &$G_a$     &$G^c_a$   &$\Phi_a$   &$\Phi^c_3$
         &$\Phi^c_i$\\
      \hline
$S_4$    &${\bf 2}$ &${\bf 1}$  &${\bf 2}$    &${\bf 1}$ &${\bf 3}$ &${\bf 3}$ &${\bf 3}$  &${\bf 1}$ 
         &${\bf 2}$ \\
      \hline
$Z^{(2)}_2$&$-$     &$+$        &$-$          &$+$       &$+$       &$+$       &$+$        &$+$
         &$+$       \\  
      \hline 
$Z^N_2$  &$+$       &$+$        &$+$          &$+$       &$+$       &$+$       &$+$        &$+$ 
         &$+$       \\  
      \hline      
$Z_{17}$ &$1/17$    &$0$        &$16/17$      &$0$       &$0   $    &$0$       &$0$        &$0$
         &$0$       \\ 
      \hline
$Z^R_2$  &$+$       &$+$        &$+$          &$+$       &$+$       &$+$       &$-$        &$-$ 
         &$-$       \\  
      \hline  
$Z^{(4)}_2$&$+$     &$+$        &$+$          &$+$       &$+$       &$+$       &$+$        &$+$
         &$+$       \\  
      \hline 
$Z^{(5)}_2$&$+$     &$+$        &$+$          &$+$       &$+$       &$+$       &$+$        &$+$
         &$+$       \\  
      \hline 
      \hline
         &$L_4$     &$\bar{L}_4$&$H^U_4$      &$\bar{H}^U_4$&$D_i$  &$L_J$     &$D^c_J$    &$\bar{L}_J$
         &$\bar{D}^c_J$ \\
      \hline      
$S_4$    &${\bf 1}$ &${\bf 1}$  &${\bf 1}$    &${\bf 1}$ &${\bf 2}$ &${\bf 2}$ &${\bf 2}$  &${\bf 2}$ 
         &${\bf 2}$ \\
      \hline 
$Z^{(2)}_2$&$+$     &$+$        &$+$          &$+$       &$+$       &$+$       &$+$        &$+$
         &$+$       \\  
      \hline 
$Z^N_2$  &$+$       &$+$        &$+$          &$+$       &$+$       &$+$       &$+$        &$+$ 
         &$+$       \\  
      \hline
$Z_{17}$ &$0$       &$4/17$     &$0$          &$4/17$    &$16/17$   &$0$       &$0$        &$1/17$
         &$1/17$    \\ 
      \hline     
$Z^R_2$  &$+$       &$+$        &$-$          &$-$       &$+$       &$+$       &$+$        &$+$ 
         &$+$       \\  
      \hline   
$Z^{(4)}_2$&$-$     &$-$        &$-$          &$-$       &$+$       &$+$       &$+$        &$+$
         &$+$       \\  
      \hline
$Z^{(5)}_2$&$+$     &$+$        &$+$          &$+$       &$+$       &$-$       &$-$        &$-$
         &$-$       \\  
      \hline         
\end{tabular}
\end{center}
\caption{$S_4\times Z^{(2)}_2\times Z^N_2\times Z_{17}\times Z^R_2\times Z^{(4)}_2\times Z^{(5)}_2$ 
assignment of superfields
(Where the indices  $i$ and $J$ of the $S_4$ doublets runs $i=1,2$ and $J=5,6$ respectively,
and the index $a$ of the $S_4$ triplets runs $a=1,2,3$.)}
\end{table}

\subsection{SUSY breaking}

For the successful leptogenesis, the symmetry
$Z^{(2)}_2\times Z^N_2$ must be broken softly.
Therefore we assume these symmetries are broken in
hidden sector and the effects are mediated to observable sectors by gravity.
We introduce hidden sector superfields $A,B_+,B_{1,-},B_{2-},C_+,C_{1,-},C_{2-}$,
where their representations are given in Table 3.

\begin{table}[htbp]
\begin{center}
\begin{tabular}{|c|c|c|c|c|c|c|c|}
\hline
           &$A$  &$B_+$&$B_{1-}$&$B_{2-}$&$C_+$&$C_{1-}$&$C_{2-}$ \\  
\hline        
$Z^{(2)}_2$&$+$  &$+$  &$-$     &$+$     &$+$  &$-$     &$+$    \\ 
\hline
$Z^N_2$    &$+$  &$+$  &$+$     &$-$     &$+$  &$+$     &$-$    \\ 
\hline
$Z^H_2$    &$+$  &$-$  &$-$     &$-$     &$-$  &$-$     &$-$    \\ 
\hline
$U(1)_R$   &$2$  &$2$  &$2$     &$2$     &$0$  &$0$     &$0$    \\
\hline
\end{tabular}
\end{center}
\caption{$Z^{(2)}_2\times Z^N_2\times Z^H_2\times U(1)_R$ assignment of hidden sector superfields.
All these superfields are trivial under the gauge symmetry $G_{32111}$ and flavor symmetry
$S_4\times Z_{17}\times Z^R_2 \times Z^{(4)}_2\times Z^{(5)}_2$.
The observable sector superfields are $Z^H_2$-even.}
\end{table}

We construct O'Raifeartaigh model by these hidden sector superfields as follow \cite{orai}
\eqn{
W_{\mbox{hidden}}&=&-M^2A+m_+B_+C_++m_{1-}B_{1-}C_{1-}
+m_{2-}B_{2-}C_{2-} \no \\
&+&\frac12\lambda_+AC^2_++\frac12\lambda_{1-} AC^2_{1-}
+\frac12\lambda_{2-} AC^2_{2-} .
}
As the F-terms of hidden sector superfields given by
\eqn{
F_A&=&-M^2+\frac12\lambda_+C^2_+ +\frac12\lambda_{1-}C^2_{1-}
+\frac12\lambda_{2-}C^2_{2-}, \\
F_{B_+}&=&m_+C_+ ,\\
F_{B_{1-}}&=&m_{1-}C_{1-} ,\\
F_{B_{2-}}&=&m_{2-}C_{2-} ,\\
F_{C_+}&=&m_+B_++\lambda_+AC_+ ,\\
F_{C_{1-}}&=&m_-B_{1-}+\lambda_{1-} AC_{1-}, \\
F_{C_{2-}}&=&m_-B_{2-}+\lambda_{2-} AC_{2-},
}
do not have the solution as
\eqn{
F_A=F_{B_+}=F_{B_{1-}}=F_{B_{2-}}=0,
}
supersymmetry is spontaneously broken.
The flavor symmetry $Z^{(2)}_2\times Z^N_2$ is also broken.

Since we assume the $U(1)_R$ symmetry is explicitly broken
in the higher dimensional terms \cite{rsym},
soft SUSY breaking terms are induced by
the interaction terms between observable sector and hidden sector as
\eqn{
{\cal L}_{SB}=\L\{\L[\frac{A}{M_P}W^\alpha W_\alpha
+\frac{A}{M_P}(c_{ABC}X_AX_BX_C+\cdots)\R]_F+h.c.\R\}
+\L[\frac{A^*A}{M^2_P}c_{ab}X^*_aX_b +h.c. \R]_D ,
}
where the indices  $A,B,C$ runs the species of superfields and
the indices $a,b,c$ runs generation numbers.
Generally, as the coefficient matrices $c_{ab}$ are not unit matrices,
large flavor changing processes are induced by the sfermion exchange.
The explicit $Z^{(2)}_2\times Z^N_2$ breaking terms are given by
\eqn{
{\cal L}_{Z^{(2)}_2B}
&=&\L[\frac{B^*_+B_{1-}}{M^2_P}\frac{(D_1X_1+D_2X_2)^*X_3}{M_P}+h.c.\R]_D \no \\
&=&\epsilon m^2_{BX}\L(c(X_1)^*X_3+s(X_2)^*X_3\R)+h.c. \quad (X=H^U,H^D,S) , \\
{\cal L}_{Z^N_2B}
&=&\L[\frac{B^*_+B_{2-}}{M^2_P}(c_2N^c_2+c_3N^c_3)^*(N^c_1)+h.c.\R]_D
=m^2_{12}(N^c_2)^*N^c_1+m^2_{13}(N^c_3)^*N^c_1+h.c. .
}

\subsection{$S_3$ breaking}

The $S_3$ subgroup of $S_4$ is broken by the VEV of $S_4$-doublet flavon $D_i$.
Here we consider the direction of VEV. For the later convenience, we define the 
products of $D_i$ as follows,
\eqn{
{\bf 1}&:&E_2=D^2_1+D^2_2,\quad E_3=3D^2_1D_2-D^3_2, \\
{\bf 1'}&:&P_3=D^3_1-3D_1D^2_2, \\
{\bf 2}&:&V_1=\2tvec{D_1}{D_2},\quad V_2=\2tvec{2D_1D_2}{D^2_1-D^2_2}, \quad
V_4=\2tvec{-D_2P_3}{D_1P_3} , \quad
V_5=\2tvec{-(D^2_1-D^2_2)P_3}{2D_1D_2P_3},
}
and the VEVs of each components of $D_i$ as
\eqn{
\L<D_1\R>=V_Dc=V_D\cos\theta,\quad \L<D_1\R>=V_Ds=V_D\sin\theta .
}
Generally,  the superpotential of $D_i$ is written in the form of
polynomial in $E_2,E_3,P_3$ as
\eqn{
M^{14}_PW_D=a_1E^7_2E_3+a_2E^4_2E^3_3+a_3E^4_2P^2_3E_3
+a_4E_2E^5_3+a_5E_2P^2_3E^3_3+a_6E_2P^4_3E_3.
}
Substituting the VEVs given in Eq.(43) to the flavon potential, we get
\eqn{
V(V_D,\theta)=\L\{-A[a'_1s_3+a'_2s^3_3+a'_3c^2_3s_3+a'_4s^5_3+a'_5c^2_3s^3_3+a'_6c^4_3s_3]
V^3_D\L(\frac{V_D}{M_P}\R)^{14}+h.c.\R\}
+V_F+m^2V^2_D,
}
where
\eqn{
s_3=\sin3\theta,\quad c_3=\cos3\theta,
}
and $V_F$ is F-term contribution. As this potential is polynomial in $s_3$,
the stationary condition
\eqn{
\frac{\pl V(V_D,\theta)}{\pl \theta}=c_3\L[a''_0+a''_1s_3+a''_2s^2_3
+a''_3s^3_3+a''_4s^4_3+a''_5s^5_3+a''_7s^7_3+ a''_9s^9_3 \R]=0,
}
gives parameter independent solution
\eqn{
c_3=0,
}
and parameter dependent solution
\eqn{
a''_0+a''_1s_3+a''_2s^2_3
+a''_3s^3_3+a''_4s^4_3+a''_5s^5_3+a''_7s^7_3+ a''_9s^9_3=0.
}
Which solution of two is selected for the global minimum
is depends on the parameters in potential. 
Since the solution Eq.(48) gives wrong prediction such as
massless up-quark and electron, we assume
the solution Eq.(49) corresponds to the global minimum. 
In this paper we assume $\L<D_i\R>$ are real without any reason,
which is important in considering CP violation in section 4.

The scale of $V_D$ is determined by the minimum condition
\eqn{
\frac{1}{V_D}\frac{\pl V(V_D,\theta)}{\pl V_D}\sim m^2+\frac{V^{30}_D}{M^{28}_P}=0,
}
as
\eqn{
\frac{V_D}{M_P}\sim \L(\frac{|m|}{M_P}\R)^{1/15}
\sim \L(\frac{10^3\mbox{GeV}}{10^{18}\mbox{GeV}}\R)^{1/15}
\sim 10^{-1},
}
which agrees with Eq.(22). In this paper we sometimes write SUSY breaking
scalar squared mass parameters as $m^2$ for simplicity and assume $m\sim O(\mbox{TeV})$.

\subsection{$S_4$ breaking}

The superpotential of gauge non-singlet
flavons $\Phi,\Phi^c$ is given by
\eqn{
W_\Phi&=&\frac{Y^\Phi_1}{M_P}(\Phi^c_3)^2[\Phi^2_1+\Phi^2_2+\Phi^2_3]
+\frac{Y^\Phi_2}{M_P}[(\Phi^c_1)^2+(\Phi^c_2)^2][\Phi^2_1+\Phi^2_2+\Phi^2_3] \no \\
&+&\frac{Y^\Phi_3}{M_P}
[2\sqrt{3}\Phi^c_1\Phi^c_2(\Phi^2_2-\Phi^2_3)
+((\Phi^c_1)^2-(\Phi^c_2)^2)(\Phi^2_2+\Phi^2_3-2\Phi^2_1)] \no \\
&+&\frac{Y^\Phi_4}{M_P}\Phi^c_3[\sqrt{3}\Phi^c_1(\Phi^2_2-\Phi^2_3)
+\Phi^c_2(\Phi^2_2+\Phi^2_3-2\Phi^2_1)]. 
}
Since the first term in Eq.(25) drives
the squared mass of $\Phi^c_3$ to be negative through RGEs,
these flavons develop VEVs along the D-flat direction as follows 
\eqn{
\L<\Phi^c_1\R>=\L<\Phi^c_2\R>=0,\quad
\L<\Phi_1\R>=\L<\Phi_2\R>=\L<\Phi_3\R>=\frac{\L<\Phi^c_3\R>}{\sqrt{3}}
=\frac{V}{\sqrt{3}},
}
where $S_3$-symmetry is unbroken in this vacuum. The
scale of $V$ is determined by the minimum condition
\eqn{
\frac{1}{V}\frac{\pl V(\Phi)}{\pl\Phi}\sim m^2+|Y^\Phi|^2\frac{V^4}{M^2_P}=0,
}
as
\eqn{
\frac{V}{M_P}\sim \sqrt{\frac{|m|}{|Y^\Phi|M_P}}
\sim \sqrt{\frac{10^3\mbox{GeV}}{(0.1)10^{18}\mbox{GeV}}}\sim 10^{-7},
}
which agrees with Eq.(16). In this paper we define the size of $O(1)$ coefficient
as $0.1<Y^X<1.0$.

Note that there are $S_3$ breaking corrections in the potential of $\Phi,\Phi^c$
as follows
\eqn{
V(\Phi)&\supset&m^2_1\L(\frac{(D_1\Phi^c_2+D_2\Phi^c_1)^*(D_1\Phi^c_3)
+(D_1\Phi^c_1-D_2\Phi^c_2)^*(D_2\Phi_3)}{M^2_P}+h.c.\R) \no \\
&+&m^2_2\L(\frac{|2D_2\Phi_1|^2+|(\sqrt{3}D_1+D_2)\Phi_2|^2
+|(\sqrt{3}D_1-D_2)\Phi_3|^2}{M^2_P}\R)+\cdots \no \\
&=&\epsilon^2m^2_1[s_2(\Phi^c_1)^*\Phi^c_3+c_2(\Phi^c_2)^*\Phi^c_3+h.c.]\no \\
&+&\epsilon^2m^2_2[4s^2|\Phi_1|^2+(\sqrt{3}c+s)^2|\Phi_2|^2
+(\sqrt{3}c-s)^2|\Phi_3|^2]+\cdots,
}
the direction given in Eq.(53) is modified as follows
\eqn{
\L<\Phi^c_1\R>\sim \L<\Phi^c_2\R>\sim O(\epsilon^2)V,\quad
\L<\Phi^c_3\R>= (1+O(\epsilon^2))V,\quad
\L<\Phi_a\R>=(1+O(\epsilon^2))\frac{V}{\sqrt{3}}.
}



\section{Higgs Sector}

Based on the set up given in section 2, we discuss  
about phenomenology of our model.
In this section, we consider Higgs doublet multiplets $H^U_a,H^D_a$
and singlet multiplets $S_a$.

\subsection{Higgs sector}

The superpotential of Higgs sector is given by
\eqn{
W_S&=&\frac{(\lambda_1)_0}{M^2_P}S_3(D^2_1+D^2_2)(H^U_1H^D_1+H^U_2H^D_2) \no \\
&+&\frac{\lambda'_1}{M^2_P}S_3(D_1H^U_1+D_2H^U_2)(D_1H^D_1+D_2H^D_2) \no \\
&+&\frac{\lambda''_1}{M^2_P}S_3(D_2H^U_1-D_1H^U_2)(D_2H^D_1-D_1H^D_2) \no \\
&+&\frac{\lambda'''_1}{M^2_P}S_3[(2D_1D_2)(H^U_1H^D_2+H^U_2H^D_1)
+(D^2_1-D^2_2)(H^U_1H^D_1-H^U_2H^D_2)] \no \\
&+&\frac{\lambda''''_1}{M^2_P}S_3
[(D_1H^U_2+D_2H^U_1)(D_1H^D_2+D_2H^D_1)+(D_1H^U_1-D_2H^U_2)(D_1H^D_1-D_2H^D_2)] \no \\
&+&\lambda_3S_3H^U_3H^D_3 
+\lambda_4H^U_3(S_1H^D_1+S_2H^D_2)+\lambda_5(S_1H^U_1+S_2H^U_2)H^D_3 \no \\
&+&kS_3(G_1G^c_1+G_2G^c_2+G_3G^c_3).
}
For simplicity we assume
\eqn{
\lambda'_1=\lambda''_1=\lambda'''_1=\lambda''''_1=0,\quad 
(\lambda_1)_0\epsilon^2=\lambda_1.
}
The coupling $k$ and $\lambda_3$ drive the squared mass of $S_3$ to be negative.

Omitting $O(\epsilon)$-terms, Higgs potential is given by
\eqn{
V&=&m^2_{H^U}(|H^U_1|^2+|H^U_2|^2)+m^2_{H^U_3}|H^U_3|^2
+m^2_{H^D}(|H^D_1|^2+|H^D_2|^2)+m^2_{H^D_3}|H^D_3|^2 \no \\
&+&m^2_S(|S_1|^2+|S_2|^2)+m^2_{S_3}|S_3|^2 \no \\
&-&\left\{ \lambda_3A_3S_3H^U_3H^D_3 
+ \lambda_4A_4H^U_3(S_1H^D_1+S_2H^D_2)+\lambda_5A_5(S_1H^U_1+S_2H^U_2)H^D_3 +h.c. \right\} \no \\
&+&\left|\lambda_3H^U_3H^D_3 \right|^2
+\left|\lambda_4H^U_3H^D_1+\lambda_5H^D_3H^U_1\right|^2
+\left|\lambda_4H^U_3H^D_2+\lambda_5H^D_3H^U_2\right|^2 \no \\
&+&\left|\lambda_3S_3H^D_3+\lambda_4(S_1H^D_1+S_2H^D_2)\right|^2
+\left|\lambda_5H^D_3S_1\right|^2+\left|\lambda_5H^D_3S_2\right|^2 \no \\
&+&\left|\lambda_3S_3H^U_3+\lambda_5(S_1H^U_1+S_2H^U_2)\right|^2
+\left|\lambda_4H^U_3S_1\right|^2+\left|\lambda_4H^U_3S_2\right|^2 \no \\
&+&\frac18g^2_2\sum^3_{A=1}\L[(H^U_a)^\dagger\sigma_AH^U_a+(H^D_a)^\dagger\sigma_AH^D_a\R]^2
+\frac18g^2_Y\L[|H^U_a|^2-|H^D_a|^2\R]^2 \no \\
&+&\frac12 g^2_x\L[-2|H^U_a|^2-3|H^D_a|^2+5|S_a|^2\R]^2+V_{\mbox{1-loop}},
}
where $V_{\mbox{1-loop}}$ is 1-loop corrections from $Q_3,U^c_3,G_a,G^c_a$.
The VEVs of $H^U_3,H^D_3,S_3$ trigger off gauge symmetry breaking at low energy scale.
$Z^{(2)}_2$-breaking terms
\eqn{
V_{FB}=\epsilon m^2_{BU}(H^U_1c+H^U_2s)^*H^U_3
+\epsilon m^2_{BU}(H^D_1c+H^D_2s)^*H^D_3
+\epsilon m^2_{BS}(S_1c+S_2s)^*S_3+h.c. ,
}
enforce $S_4$-doublets developing VEVs as follows
\eqn{
\L<X_1\R>\sim c\L(\frac{\epsilon m^2_{BX}}{m^2_X}\R)\L<X_3\R>,\quad
\L<X_2\R>\sim s\L(\frac{\epsilon m^2_{BX}}{m^2_X}\R)\L<X_3\R>,\quad
X=H^U,H^D,S.
}
Due to the $Z^{(2)}_2$ symmetry, the Yukawa couplings between $H^U_3$ and $N^c$ are forbidden
and neutrino Dirac mass is not induced. To give neutrino Dirac mass,
we assume the size of VEV of $H^U_i$ is given by
\eqn{
\L<H^U_{1,2}\R>\sim \epsilon^2\L<H^U_3\R> \sim 1\mbox{GeV},
}
and put the $Z^{(2)}_2$ breaking parameters as follows
\eqn{
m^2_{BU}\sim m^2_{BD}\sim m^2_{BS}\sim \epsilon m^2_{\mbox{SUSY}},
}
by hand. The suppression factor $O(\epsilon)$ may be induced 
by the running based on RGEs, 
because off diagonal elements of scalar squared mass matrix do not receive
the contributions from gaugino mass parameters
which tend to make scalar squared mass  larger at low energy scale.

We use the notation of VEVs as follows
\eqn{
\L<H^U_i\R>=(c,s)v_u ,\quad \L<H^U_3\R>=v'_u, \quad
\L<H^D_i\R>=(c,s)v_d ,\quad \L<H^D_3\R>=v'_d, \quad  
\L<S_i\R>=(c,s)v_s   ,\quad \L<S_3\R>=v'_s,
}
where we fix the values by
\eqn{
v'_u=150.7,\quad v'_d=87.0,\quad v'_s=4000,\
v_{EW}=\sqrt{(v'_u)^2+(v'_d)^2}=174 \  (\mbox{GeV}) ,\
\tan\beta=\frac{v'_u}{v'_d}=\tan\frac{\pi}{3}.
}
In this paper, we neglect the contributions from $v_{u,d,s}$ except for neutrino sector.
With this approximation, the potential minimum conditions are given as follows,
\eqn{
0=\frac{1}{v'_u}\frac{\partial V}{\partial H^U_3}
&=&m^2_{H^U_3}-\lambda_3A_3v'_s(v'_d/v'_u)+\lambda^2_3(v'_d)^2
+\lambda^2_3(v'_s)^2  +\frac14(g^2_Y+g^2_2)[(v'_u)^2-(v'_d)^2] \no \\
&-&2g^2_x[-2(v'_u)^2-3(v'_d)^2+5(v'_s)^2] 
+\frac{1}{2v'_u}\frac{\partial V_{1-\mbox{loop}}}{\partial v'_u},  \\
0=\frac{1}{v'_d}\frac{\partial V}{\partial H^D_3}
&=&m^2_{H^D_3}-\lambda_3A_3v'_s(v'_u/v'_d)+\lambda^2_3(v'_u)^2 
+\lambda^2_3(v'_s)^2-\frac14(g^2_Y+g^2_2)[(v'_u)^2-(v'_d)^2] \no \\
&-&3g^2_x[-2(v'_u)^2-3(v'_d)^2+5(v'_s)^2] 
+\frac{1}{2v'_d}\frac{\partial V_{1-\mbox{loop}}}{\partial v'_d}, \\
0=\frac{1}{v'_s}\frac{\partial V}{\partial S_3}
&=&m^2_{S_3}-\lambda_3A_3v'_u(v'_d/v'_s)
+\lambda^2_3(v'_u)^2+\lambda^2_3(v'_d)^2
+5g^2_x[-2(v'_u)^2-3(v'_d)^2+5(v'_s)^2], 
}
where the 1-loop contribution is neglected in Eq.(69),
which is unimportant. These equations give the boundary conditions
for $m^2_{H^U_3},m^2_{H^D_3},m^2_{S_3}$ at SUSY breaking scale
$M_S=10^3\mbox{GeV}$ in solving RGEs.

The mass matrices of heavy Higgs bosons are given as follows
\eqn{
M^2_3(\mbox{CP even})&\simeq&\Mat3{\lambda_3A_3v'_sv'_d/v'_u}{-\lambda_3A_3v'_s}{0}
{-\lambda_3A_3v'_s}{\lambda_3A_3v'_sv'_u/v'_d}{0}
{0}{0}{50g^2_x(v'_s)^2} \\
M^2_i(\mbox{CP even})&=&M^2_i(\mbox{CP odd})
\simeq \mbox{diag}\L(m^2_{H^U}-10g^2_x(v'_s)^2,
m^2_{H^D}-15g^2_x(v'_s)^2,
m^2_S+25g^2_x(v'_s)^2\R) \\
M^2_3(\mbox{CP-odd})&\simeq&\lambda_3A_3v'_s\Mat3{v'_d/v'_u}{1}{0}
{1}{v'_u/v'_d}{0}
{0}{0}{0} \\
M^2_3(\mbox{charged})&\simeq&\lambda_3A_3v'_s\mat2{v'_d/v'_u}{1}{1}{v'_u/v'_d} \\
M^2_i(\mbox{charged})&\simeq&\mbox{diag}
\L(m^2_{H^U}-10g^2_x(v'_s)^2, 
m^2_{H^D}-15g^2_x(v'_s)^2\R),
}
where only $O(1\mbox{TeV})$ terms are considered and notation  
$H^UH^D=(H^U)^0(H^D)^0-(H^U)^+(H^D)^-$ is used. 
In this approximation, generation mixing terms are negligible.
The third generation mass matrices are diagonalized as follows
\eqn{
M^2_3(\mbox{CP even})&=&\mbox{diag}\L(0,\frac{2\lambda_3A_3v'_s}{\sin2\beta} ,
50g^2_x(v'_s)^2\R), \\
M^2_3(\mbox{CP odd})&=&\mbox{diag}\L(0,\frac{2\lambda_3A_3v'_s}{\sin2\beta} ,0\R), \\
M^2_3(\mbox{charged})&=&\mbox{diag}\L(0,\frac{2\lambda_3A_3v'_s}{\sin2\beta}\R),
}
where the zero eigenvalue in CP even Higgs bosons corresponds to
lightest neutral CP even Higgs boson and the other zero eigenvalues are
Nambu-Goldstone modes absorbed into gauge bosons.

\subsection{Lightest neutral CP even Higgs boson}

To calculate the mass of the lightest neutral CP even Higgs boson,
we diagonalize $2\times 2$ sub-matrix given by
\eqn{
M^2_3(\mbox{even})&=&\mat2{m^2_{uu,3}}{m^2_{ud,3}}
{m^2_{ud,3}}{m^2_{dd,3}} , \\
m^2_{uu,3}&=&\lambda_3A_3(v'_sv'_d/v'_u)
+\frac12(g^2_Y+g^2_2)(v'_u)^2+8g^2_x(v'_u)^2
+\frac12\frac{\partial^2 V_{1-\mbox{loop}}}{\partial (v'_u)^2}
-\frac{1}{2v'_u}\frac{\partial V_{1-\mbox{loop}}}{\partial v'_u}, \\
m^2_{ud,3}&=&-\lambda_3A_3v'_s+2\lambda^2_3v'_uv'_d 
-\frac12 (g^2_Y+g^2_2)v'_uv'_d+12g^2_xv'_uv'_d 
+\frac12\frac{\partial^2 V_{1-\mbox{loop}}}{\pl v'_u\pl v'_d}, \\
m^2_{dd,3}&=&\lambda_3A_3(v'_sv'_u/v'_d)+\frac12(g^2_Y+g^2_2)(v'_d)^2+18g^2_x(v'_d)^2
+\frac12\frac{\partial^2 V_{1-\mbox{loop}}}{\partial (v'_d)^2}
-\frac{1}{2v'_d}\frac{\partial V_{1-\mbox{loop}}}{\partial v'_d},
}
where $O(v_{EW})$ terms are included. 
We evaluate the 1-loop contributions from top, stop, G-Higgs and
G-higgsino as
\eqn{
V_{1-\mbox{loop}}
=\frac{1}{64\pi^2}\mbox{Str}\L[M^4\L(\ln\frac{M^2}{\Lambda^2}-\frac32\R)\R],
}
at the renormalization point $\Lambda=M_S$ \cite{1-loop}\cite{u1-1-loop}, where
the mass eigenvalues are given by
\eqn{
\begin{array}{ll}
m^2_{T_\pm}=M^2_T+(Y^U_3v'_u)^2\pm R_T , &
m^2_{G_{\pm}}=M^2_G\pm R_G , \\
M^2_T=\frac12\L(m^2_{Q_3}+m^2_{U_3}\R)+5g^2_x(v'_s)^2 , &
M^2_G=\frac12\L(m^2_G+m^2_{G^c}\R)+(kv'_s)^2-\frac{25}{2}g^2_x(v'_s)^2 , \\
R_T=\sqrt{(\Delta M^2_T)^2+(Y^U_3X_T)^2}  , &
R_G=\sqrt{(\Delta M^2_G)^2+(kX_G)^2} , \\
\Delta M^2_T= \frac12\L(m^2_{Q_3}-m^2_{U_3}\R) , &
\Delta M^2_G=\frac12\L(m^2_G-m^2_{G^c}\R)+\frac{5}{2}g^2_x(v'_s)^2 , \\
X_T=\lambda_3v'_sv'_d-A^U_3v'_u  , &
X_G=\lambda_3v'_uv'_d-A_kv'_s , \\
m_t=Y^U_3v'_u  ,&
m_g=kv'_s, \\
\end{array}
}
where we neglect  $O(v_{EW})$ terms in D-term contributions.
The mass matrices of stop and G-Higgs  are given in following sections (see Eq.(171) and Eq.(236)).
By the rotation with
\eqn{
V_H=\frac{1}{v_{EW}}\mat2{v'_u}{-v'_d}{v'_d}{v'_u},
}
the $O(1\mbox{TeV})$ term is eliminated from off-diagonal element of Eq.(78) and
we get
\eqn{
m^2_h&=&(V^T_HM^2V_H)_{11}=(m^2_h)_0+(m^2_h)_T+(m^2_h)_G , \\
(m^2_h)_0&=&\L[(\lambda_3\sin\beta)^2+\frac{g^2_Y+g^2_2}{2}\cos^22\beta
+2g^2_x(2\sin^2\beta+3\cos^2\beta)^2\R]v^2_{EW} , \\
(m^2_h)_T&=&\frac{3(Y^U_3)^2}{16\pi^2v^2_{EW}}
\L[
\frac{(Y^U_3X^2_T)^2}{R^2_T}
+2(Y^U_3(v'_u)^2)^2\L(\ln\frac{m^2_{T_+}}{(Y^U_3v'_u)^2}
+\ln\frac{m^2_{T_-}}{(Y^U_3v'_u)^2}\R)  \R. \no \\
&+&\L. \L(\frac{2(Y^U_3v'_u)^2X^2_T}{R_T}
-\L[M^2_T+(Y^U_3v'_u)^2\R]\frac{(Y^U_3X^2_T)^2}{2R^3_T}
\R)\ln\frac{m^2_{T_+}}{m^2_{T_-}}\R] , \\
(m^2_h)_G&=&\frac{9k^2(\lambda_3v'_uv'_d)^2}{8\pi^2v^2_{EW}}
\L[-2\frac{(\Delta M^2_G)^2}{R^2_G}
+\frac{M^2_G(\Delta M^2_G)^2}{R^3_G}\ln\frac{m^2_{G_+}}{m^2_{G_-}}
+\ln\frac{m^2_{G_+}}{\Lambda^2}+\ln\frac{m^2_{G_-}}{\Lambda^2}\R].
}
If we fix the parameters at $M_S$ as given in Table 6, we get
\eqn{
m_h=125.7 ,\quad 
\sqrt{(m^2_h)_0}=82.7 ,\quad 
\sqrt{(m^2_h)_T}=94.2 ,\quad
\sqrt{(m^2_h)_G}=9.2 \quad (\mbox{GeV}).
}
The 1-loop contribution is dominated by stop and top contributions,
this is because we put $k$ small ($k=0.5$) to intend the mass values 
of the particles in the loops are within
the testable range of LHC at $\sqrt{s}=14\mbox{TeV}$ as follows
\eqn{
m_{T_+}=1882,\quad m_{T_-}=1178,\quad m_{G_+}=3908,\quad
m_{G_-}=1737,\quad m_g=2000 \quad (\mbox{GeV}).
}
The value of $\lambda_3=0.37$ is tuned to realize observed Higgs mass
which is mainly controlled by this parameter through $(\lambda_3v_{EW}\sin\beta)^2$
and $X_T$ for fixed $v'_s$ and $A^U_3(=A_t)$.

\subsection{Chargino and neutralino}

At next we consider the higgsinos and the singlinos.
The mass matrix of the charged higgsinos is given by
\eqn{
{\cal L}_C&=&((h^U_1)^+,(h^U_2)^+,(h^U_3)^+)
\Mat3{\lambda_1v'_s}{0}{\lambda_5v_sc}
{0}{\lambda_1v'_s}{\lambda_5v_ss}
{\lambda_4v_sc}{\lambda_4v_ss}{\lambda_3v'_s}
\3tvec{(h^D_1)^-}{(h^D_2)^-}{(h^D_3)^-} +h.c..
}
Since the (3,3) element is much larger than the other $O(\epsilon^2)$ elements,
the first and second generation higgsinos decouples and have the
same mass $\lambda_1v'_s$.
With the gaugino interaction given as follows
\eqn{
{\cal L}_{\mbox{gaugino}}&=&-i\sqrt{2}(H^U_a)^\dagger
\L[g_2\sum^3_{A=1}\lambda^A_2T^A_2+\frac12 g_Y\lambda_Y-2g_x\lambda_X\R]h^U_a \no \\
&-&i\sqrt{2}(H^D_a)^\dagger 
\L[g_2\sum^3_{A=1}\lambda^A_2T^A_2-\frac12 g_Y\lambda_Y-3g_x\lambda_X\R]h^D_a
-i\sqrt{2}(S_a)^\dagger[5g_x\lambda_X]s_a \no \\
&-&\frac12M_2\lambda^A_2\lambda^A_2
-\frac12M_Y\lambda_Y\lambda_Y
-\frac12M_X\lambda_X\lambda_X+h.c. ,
}
the third generation charged higgsino mixes with wino and 
the mass matrix is given by
\eqn{
{\cal L}&\supset&((h^U_3)^+,w^+)
\mat2{\lambda_3v'_s}{g_2v'_u}{g_2v'_d}{M_2}
\2tvec{(h^D_3)^-}{w^-} +h.c. , \\
&&w^\pm=\frac{-i}{\sqrt{2}}(\lambda^1_2\mp i\lambda^2_2).
}
The mass eigenvalues of charginos are given by
\eqn{
M\2tvec{\chi^\pm_3}{\chi^\pm_w}
&=&\frac12\L[(\lambda_3v'_s)^2+g^2_2(v'_u)^2+g^2_2(v'_d)^2+M^2_2\R] \no \\
&\pm& \sqrt{\frac14[(\lambda_3v'_s)^2+(g_2v'_u)^2-(g_2v'_d)^2-M^2_2]^2
+(\lambda_3g_2v'_sv'_d+M_2g_2v'_u)^2} , \\
M(\chi^\pm_i)&=&\lambda_1v'_s ,
}
where $\chi^\pm_3$ is almost third generation higgsino
and $\chi^\pm_w$ is almost wino.

The mass matrix of the neutralinos is divided into
two $3\times 3$ matrices and one $6\times 6$ matrix as follows
\eqn{
{\cal L}&\supset&-\frac12\sum_{i=1,2}((h^U_i)^0,(h^D_i)^0,s_i)
\Mat3{0}{\lambda_1v'_s}{\lambda_5v'_d}
{\lambda_1v'_s}{0}{\lambda_4v'_u}
{\lambda_5v'_d}{\lambda_4v'_u}{0}
\3tvec{(h^U_i)^0}{(h^D_i)^0}{s_i}
-\frac12\chi^T_0M_\chi\chi_0, \\
M_\chi&=&\L(
\begin{array}{cccccc}
0               &\lambda_3v'_s    &\lambda_3v'_d &g_Yv'_u/\sqrt{2}&-g_2v'_u/\sqrt{2}&-2g_xv'_u\sqrt{2}\\
\lambda_3v'_s   &0                &\lambda_3v'_u &-g_Yv'_d/\sqrt{2}&g_2v'_d/\sqrt{2}&-3g_xv'_d\sqrt{2}\\
\lambda_3v'_d   &\lambda_3v'_u    &0             &0                &0               &5g_xv'_s\sqrt{2}\\
g_Yv'_u/\sqrt{2}&-g_Yv'_d/\sqrt{2}&0             & -M_Y            &0               &0    \\
-g_2v'_u/\sqrt{2}&g_2v'_d/\sqrt{2}&0             &0                &-M_2            &0    \\
-2g_xv'_u\sqrt{2}&-3g_xv'_d\sqrt{2}&5g_xv'_s\sqrt{2}&0             &0               &-M_X \\
\end{array}
\R) , \\
\chi^T_0&=&((h^U_3)^0,(h^D_3)^0,s_3,i\lambda_Y,i\lambda^3_2,i\lambda_X) .
}
The mass eigenvalues of these mass matrices are given in Table 7.
The common mass of two LSPs is given by the smallest eigenvalue of $3\times 3$ matrix given in Eq.(97).
Note that the LEP bound for chargino \cite{lep2}
\eqn{
\lambda_1v'_s> 100\mbox{GeV} , 
}
must be satisfied. Requiring the coupling constants $\lambda_{4,5}$ do not
blow up in $\mu<M_P$, we put upper bound for them as $\lambda_4<0.5,\lambda_5<0.7$,
then the rough estimation of LSP mass is given by
\eqn{
M(\chi^0_{i,1})\sim \frac{2(\lambda_4v'_u)(\lambda_5v'_d)}{\lambda_1v'_s}
<\frac{9000\mbox{GeV}^2}{\lambda_1v'_s},
}
where $\chi^0_{i,1}$ is almost singlino.
To realize density parameter of dark matter $\Omega_{CDM}h^2=0.11$,
we must tune $M(\chi^0_{i,1})\sim 30-35\mbox{GeV}$ to enhance annihilation cross section.
This condition gives upper bound as
\eqn{
\lambda_1v'_s< 300\mbox{GeV}.
}
This constraint is not consistent with the lower bound  from ATLAS \cite{atlas} and CMS \cite{cms} as follows
\eqn{
m_{\chi^\pm_1}>295 \mbox{GeV} , \quad
m_{\chi^\pm_1}>330 \mbox{GeV} .
}
Therefore we assume the lightest chargino mass is in the region
\eqn{
100<\lambda_1v'_s<140 \quad (\mbox{GeV}),
}
in which 3-lepton emission is suppressed due to the small mass difference
between chargino and neutralino compared with $m_Z$.

Note that bino-like neutralino can decay into Higgs boson and LSP through the
$O(\epsilon^2)$ mixing of Higgs bosons by the interaction
\eqn{
{\cal L}\supset -i\sqrt{2}(H^U_1)^0\L(\frac12g_Y\lambda_Y\R)(h^U_1)^0
\sim O(\epsilon^2)(H^U_3)^0\lambda_Y(h^U_1)^0.
}



\section{Quark and Lepton  Sector}

In this section, we consider the quark and lepton sector and
test our model by observed values given as follows,
running masses of quarks and charged leptons at $\mu=M_S=1\mbox{TeV}$ \cite{mass}
\eqn{
\begin{tabular}{lll}
$m_u=1.10^{+0.43}_{-0.37} (\mbox{MeV})$, &
$m_c=532\pm 74 (\mbox{MeV})$, &
$m_t=150.7\pm 3.4 (\mbox{GeV})$, \\
$m_d=2.50^{+1.08}_{-1.03} (\mbox{MeV})$, &
$m_s=47^{+14}_{-13} (\mbox{MeV})$, &
$m_b=2.43\pm 0.08 (\mbox{GeV})$, \\
$m_e=0.4959 (\mbox{MeV})$, &
$m_\mu=104.7 (\mbox{MeV})$, &
$m_\tau=1780 (\mbox{MeV})$,
\end{tabular}
}
CKM matrix elements \cite{PDG2012}
\eqn{
&&
\begin{tabular}{lll}
$\L| V_{ud}\R|=0.97427$, &
$\L| V_{us}\R|=0.22534$,   &
$\L| V_{ub}\R|=0.00351$, \\
$\L| V_{cd}\R|=0.22520$, &
$\L| V_{cs}\R|=0.97344$,   &
$\L| V_{cb}\R|=0.0412$, \\
$\L| V_{td}\R|=0.00867$, &
$\L| V_{ts}\R|=0.0404$,&
$\L| V_{tb}\R|=0.999146$,
\end{tabular} 
}
and neutrino masses and MNS mixing angles \cite{PDG2012}
\begin{eqnarray}
\Delta m^2_{21}&=& m^2_{\nu_2}-m^2_{\nu_1}
=(7.58^{+0.22}_{-0.26})\times 10^{-5} \quad (\mbox{eV}^2),  \\
\Delta m^2_{32} &=& \L|m^2_{\nu_3}-m^2_{\nu_2}\R|
=(2.35^{+0.12}_{-0.09})\times 10^{-3}\quad (\mbox{eV}^2),  \\
V_{MNS}&=&\Mat3{c_{12}c_{13}}{s_{12}c_{13}}{s_{13}e^{-i\delta}}
{-s_{12}c_{23}-c_{12}s_{23}s_{13}e^{i\delta}}{c_{12}c_{23}-s_{12}s_{23}s_{13}e^{i\delta}}{s_{23}c_{13}}
{s_{12}s_{23}-c_{12}c_{23}s_{13}e^{i\delta}}{-c_{12}s_{23}-s_{12}c_{23}s_{13}e^{i\delta}}{c_{23}c_{13}}
\Mat3{e^{i\alpha_1/2}}{0}{0}{0}{e^{i\alpha_2/2}}{0}{0}{0}{1}, \no \\
&&\sin^2\theta_{12}=0.306^{+0.018}_{-0.015}, \quad
\sin^2\theta_{23}=0.42^{+0.08}_{-0.03}, \quad
\sin^2\theta_{13}=0.021^{+0.007}_{-0.008}.
\end{eqnarray}
After that we estimate the flavor changing process induced by sfermion exchange.

\subsection{Quark sector}

The superpotential of up-type quark sector is given by
\eqn{
W_U&=&Y^U_3H^U_3Q_3U^c_3+\epsilon^2Y^U_2H^U_3[Q_1s_2+Q_2c_2]U^c_3
+\epsilon^3Y^U_4H^U_3[Q_1c+Q_2s]U^c_2
+\epsilon^4Y^U_1H^U_3Q_3U^c_1  \no \\
&+&\epsilon^6\L\{Y^U_5H^U_3[Q_1s_2+Q_2c_2]U^c_1
+Y^U_6H^U_3[s_3(Q_1c+Q_2s)]U^c_1-Y^U_7H^U_3[c_3(Q_1s-Q_2c)]U^c_1\R\},
}
from which we get up-type quark mass matrix as
\eqn{
M_u=\Mat3{\epsilon^6(Y^U_5s_2+Y^U_6s_3c-Y^U_7c_3s)}{\epsilon^3Y^U_4c}{\epsilon^2Y^U_2s_2}
{\epsilon^6(Y^U_5c_2+Y^U_6s_3s+Y^U_7c_3c)}{\epsilon^3Y^U_4s}{\epsilon^2Y^U_2c_2}
{\epsilon^4Y^U_1}{0}{Y^U_3}v'_u
=\Mat3{\epsilon^6}{\epsilon^3}{\epsilon^2}
{\epsilon^6}{\epsilon^3}{\epsilon^2}
{\epsilon^4}{0}{1}v'_u .
}
Note that there is dangerous VEV direction such as $\theta=\frac{\pi}{6}$.
In this direction the matrix given in Eq.(112) is given by
\eqn{
M_u=\Mat3{\epsilon^6(Y^U_5+Y^U_6)c}{\epsilon^3Y^U_4c}{\epsilon^2Y^U_2c}
{\epsilon^6(Y^U_5+Y^U_6)s}{\epsilon^3Y^U_4s}{\epsilon^2Y^U_2s}
{\epsilon^4Y^U_1}{0}{Y^U_3}v'_u,
}
which has zero eigenvalue. 
In the same way, the down type quark masses are given by
\eqn{
W_D&=&\epsilon^4\L\{ Y^D_2H^D_3[s_3(Q_1c+Q_2s)]D^c_3
+Y^D_4H^D_3(Q_1s_2+Q_2c_2)D^c_3
+Y^D_9H^D_3[c_3(-Q_1s+Q_2c)]D^c_3\R.  \no \\
&+&\L. Y^D_5H^D_3[s_3(-Q_1s+Q_2c)]D^c_2
+Y^D_6H^D_3(-Q_1c_2+Q_2s_2)D^c_2
+Y^D_{10}H^D_3[c_3(Q_1c+Q_2s)]D^c_2\R\} \no \\
&+&\epsilon^5\L\{Y^D_8H^D_3[Q_1c+Q_2s]D^c_1
+Y^D_7H^D_3[s_3(Q_1s_2+Q_2c_2)]D^c_1 \R. \no \\
&+&\L. Y^D_{11}H^D_3[c_3(-Q_1c_2+Q_2s_2)]D^c_1\R\} 
+\epsilon^2Y^D_3H^D_3Q_3D^c_3+\epsilon^3Y^D_1s_3H^D_3Q_3D^c_1,
}
from which we get down-type quark mass matrix as follows
\eqn{
&&M_d/v'_d \no \\
&=&\Mat3{\epsilon^5(Y^D_7s_3s_2+Y^D_8c-Y^D_{11}c_3c_2)}
{\epsilon^4(-Y^D_5s_3s-Y^D_6c_2+Y^D_{10}c_3c)}
{\epsilon^4(Y^D_2s_3c+Y^D_4s_2-Y^D_9c_3s)}
{\epsilon^5(Y^D_7s_3c_2+Y^D_8s+Y^D_{11}c_3s_2)}
{\epsilon^4(Y^D_5s_3c+Y^D_6s_2+Y^D_{10}c_3s)}
{\epsilon^4(Y^D_2s_3s+Y^D_4c_2+Y^D_9c_3c)}
{\epsilon^4Y^D_1}{0}{\epsilon^2Y^D_3} \no \\
&=&\Mat3{\epsilon^5}{\epsilon^4}{\epsilon^4}
{\epsilon^5}{\epsilon^4}{\epsilon^4}
{\epsilon^4}{0}{\epsilon^2}.
}
The effects of flavor violation appear not only in superpotential
but also in K\"ahler potential as follows
\eqn{
K(U^c)&=&|U^c_1|^2+|U^c_2|^2+|U^c_3|^2
+\L\{\frac{(E_3U^c_1)^*U^c_2}{M^3_P}
+\frac{(E^2_2U^c_1)^*U^c_3}{M^4_P}
+\frac{(E_3U^c_2)^*(E_2U^c_3)}{M^5_P}+h.c.\R\} \no \\
&=&((U^c_1)^*,(U^c_2)^*,(U^c_3)^*)
\Mat3{1}{\epsilon^3}{\epsilon^4}
{\epsilon^3}{1}{\epsilon^5}
{\epsilon^4}{\epsilon^5}{1}
\3tvec{U^c_1}{U^c_2}{U^c_3}, \\
K(D^c)&=&|D^c_1|^2+|D^c_2|^2+|D^c_3|^2
+\L\{\frac{(V_2D^c_1)^*\cdot(V_1D^c_2)}{M^3_P}
+\frac{(V_2D^c_1)^*\cdot(V_1D^c_3)}{M^3_P}
+\frac{(E_3D^c_2)^*(P_3D^c_3)}{M^6_P}+h.c.\R\} \no \\
&=&((D^c_1)^*,(D^c_2)^*,(D^c_3)^*)
\Mat3{1}{\epsilon^3}{\epsilon^3}
{\epsilon^3}{1}{\epsilon^6}
{\epsilon^3}{\epsilon^6}{1}
\3tvec{D^c_1}{D^c_2}{D^c_3} , \\
K(Q)&=&(|Q_1|^2+|Q_2|^2)+|Q_3|^2 +\L\{\frac{(V_2\cdot Q)^*Q_3}{M^2_P}
+\frac{|V_1\cdot Q|^2}{M^2_P} +\cdots+ h.c. \R\} \no \\
&=&(Q^*_1,Q^*_2,Q^*_3)
\Mat3{1}{\epsilon^2}{\epsilon^2}
{\epsilon^2}{1}{\epsilon^2}
{\epsilon^2}{\epsilon^2}{1}
\3tvec{Q_1}{Q_2}{Q_3},
}
where dot in $X\cdot Y$ means inner product of two $S_4$-doublets $X,Y$.
Therefore, a superfield redefinition has to be performed in order to
get canonical kinetic terms as follows \cite{kahler}
\eqn{
\3tvec{U^c_1}{U^c_2}{U^c_3}&=&V_K(U)
\3tvec{(U^c_1)'}{(U^c_2)'}{(U^c_3)'},\quad
V_K(U)=\Mat3{1}{\epsilon^3}{\epsilon^4}
{\epsilon^3}{1}{\epsilon^5}
{\epsilon^4}{\epsilon^5}{1} , \\
\3tvec{D^c_1}{D^c_2}{D^c_3}&=&
V_K(D)
\3tvec{(D^c_1)'}{(D^c_2)'}{(D^c_3)'} , \quad
V_K(D)=
\Mat3{1}{\epsilon^3}{\epsilon^3}
{\epsilon^3}{1}{\epsilon^6}
{\epsilon^3}{\epsilon^6}{1} , \\
\3tvec{Q_1}{Q_2}{Q_3}&=&
V_K(Q)
\3tvec{Q'_1}{Q'_2}{Q'_3} ,\quad
V_K(Q)=
\Mat3{1}{\epsilon^2}{\epsilon^2}
{\epsilon^2}{1}{\epsilon^2}
{\epsilon^2}{\epsilon^2}{1},
}
by which the mass matrices given above are transformed into
\eqn{
M'_u&=&V^T_K(Q)
\Mat3{\epsilon^6}{\epsilon^3}{\epsilon^2}
{\epsilon^6}{\epsilon^3}{\epsilon^2}
{\epsilon^4}{0}{1}v'_u
V_K(U)=
\Mat3{\epsilon^6}{\epsilon^3}{\epsilon^2}
{\epsilon^6}{\epsilon^3}{\epsilon^2}
{\epsilon^4}{\epsilon^5}{1}v'_u , \\
M'_d&=&V^T_K(Q)
\Mat3{\epsilon^5}{\epsilon^4}{\epsilon^4}
{\epsilon^5}{\epsilon^4}{\epsilon^4}
{\epsilon^3}{0}{\epsilon^2}v'_d
V_K(D)=
\Mat3{\epsilon^5}{\epsilon^4}{\epsilon^4}
{\epsilon^5}{\epsilon^4}{\epsilon^4}
{\epsilon^3}{\epsilon^6}{\epsilon^2}v'_d.
}
These matrices are diagonalized as follows
\eqn{
M'_u&=&L_uM^{\mbox{diag}}_uR^\dagger_u
=\Mat3{1}{1}{\epsilon^2}{1}{1}{\epsilon^2}{\epsilon^2}{\epsilon^2}{1}
\Mat3{\epsilon^6}{0}{0}{0}{\epsilon^3}{0}{0}{0}{1}v'_u
\Mat3{1}{\epsilon^3}{\epsilon^4}
{\epsilon^3}{1}{\epsilon^5}
{\epsilon^4}{\epsilon^5}{1} , \\
M'_d&=&L_dM^{\mbox{diag}}_dR^\dagger_d
=\Mat3{1}{1}{\epsilon^2}{1}{1}{\epsilon^2}{\epsilon^2}{\epsilon^2}{1}
\Mat3{\epsilon^5}{0}{0}{0}{\epsilon^4}{0}{0}{0}{\epsilon^2}v'_d
\Mat3{1}{\epsilon^3}{\epsilon}
{\epsilon^3}{1}{\epsilon^4}
{\epsilon}{\epsilon^4}{1} .
}
Therefore Yukawa hierarchies are given by
\eqn{
&&Y_u(M_P)=\epsilon^6,\quad Y_c(M_P)=\epsilon^3,\quad Y_t(M_P)=Y^U_3(M_P)=1, \no \\
&&Y_d(M_P)=\epsilon^5,\quad Y_s(M_P)=\epsilon^4,\quad Y_b(M_P)=\epsilon^2.
}
On the other hand, observed values Eq.(106) give
\eqn{
Y_u(M_P)&=&\frac{1}{5.1}\L(\frac{1.10\times 10^{-3}}{150.7}\R)=1.4 \epsilon^6 ,\\
Y_c(M_P)&=&\frac{1}{5.1}\L(\frac{532\times 10^{-3}}{150.7}\R)=0.69 \epsilon^3 ,\\
Y_t(M_P)&=&0.28 ,\\
Y_d(M_P)&=&\frac{1}{7.2}\L(\frac{2.50\times 10^{-3}}{87}\R)=0.40\epsilon^5 ,\\
Y_s(M_P)&=&\frac{1}{7.2}\L(\frac{47\times 10^{-3}}{87}\R)=0.75\epsilon^4 ,\\
Y_d(M_P)&=&\frac{1}{7.2}\L(\frac{2.43}{87}\R)=0.39\epsilon^2 ,
}
which give good agreement with Eq.(126). Where the renormalization factors
\eqn{
\sqrt{\frac{\alpha_u(M_S)}{\alpha_u(M_P)}}=5.1,\quad
\sqrt{\frac{\alpha_d(M_S)}{\alpha_d(M_P)}}=7.2,
}
and $Y_t(M_P)=0.28$ are calculated based on RGEs given in appendix A.
CKM matrix is given by
\eqn{
V_{CKM}=L^\dagger_uL_d
=\Mat3{O(1)}{O(1)}{\epsilon^2}{O(1)}{O(1)}{\epsilon^2}{\epsilon^2}{\epsilon^2}{1},
}
which requires accidental cancellation of two mixing matrices $L_{u,d}$
to reproduce the small Cabbibo angle of CKM matrix given in Eq.(107).

Note that the $Z^{(2)}_2$ breaking induces generation mixing in Higgs bosons
then Yukawa interactions are modified as follows
\eqn{
-{\cal L}=Y^U_{ij}(H^U_3+\epsilon^2H^U_1+\epsilon^2H^U_2)q_iu^c_j
+Y^D_{ij}(H^D_3+\epsilon^2H^D_1+\epsilon^2H^D_2)q_id^c_j.
}
Since these Yukawa coupling matrices are diagonalized 
in the basis that the quark mass matrices are diagonalized,
the extra Higgs boson exchange do not contribute to the flavor changing processes.

\subsection{Lepton sector}

With the straightforward calculation, 
the mass matrices of lepton sector are given as follows. 
From the superpotentials
\eqn{
W_E&=&H^D_3(L_1,L_2,L_3)\Mat3{\epsilon^5(Y^E_7c+Y^E_8s_3s_2-Y^E_{10}c_3c_2)}
{\epsilon^3Y^E_5c}
{\epsilon^2Y^E_4s_2}
{\epsilon^5(Y^E_7s+Y^E_8s_3c_2+Y^E_{10}c_3s_2)}
{\epsilon^3Y^E_5s}
{\epsilon^2Y^E_4c_2}
{\epsilon^5Y^E_1s_3}{\epsilon^3Y^E_2s_3}{\epsilon^2Y^E_3}\3tvec{E^c_1}{E^c_2}{E^c_3} , \\
W_N&=&\epsilon^3H^U_1(L_1,L_2,L_3)
\Mat3{0}{Y^N_1s_3+Y^N_4cs_2+\cdots}{Y^N_5s_3+Y^N_8cs_2+\cdots}
{0}{-Y^N_2c_3+Y^N_4cc_2+\cdots}{-Y^N_6c_3+Y^N_8cc_2+\cdots}
{0}{Y^N_3c}{Y^N_7c}
\3tvec{N^c_1}{N^c_2}{N^c_3} \no \\
&+&\epsilon^3H^U_2(L_1,L_2,L_3)
\Mat3{0}{Y^N_2c_3+Y^N_4ss_2+\cdots}{Y^N_6c_3+Y^N_8ss_2+\cdots}
{0}{Y^N_1s_3+Y^N_4sc_2+\cdots}{Y^N_5s_3+Y^N_8sc_2+\cdots}
{0}{Y^N_3s}{Y^N_7s}
\3tvec{N^c_1}{N^c_2}{N^c_3} , \\
W_R&=&\frac{1}{M_P}(\Phi^2_1+\Phi^2_2+\Phi^2_3)
\L[Y^N_{11}N^c_1N^c_1+Y^N_{22}N^c_2N^c_2+Y^N_{33}N^c_3N^c_3
+Y^N_{23}N^c_2N^c_3\R],
}
we get original mass matrices as follows
\eqn{
M_e=\Mat3{\epsilon^5}{\epsilon^3}{\epsilon^2}
{\epsilon^5}{\epsilon^3}{\epsilon^2}
{\epsilon^5}{\epsilon^3}{\epsilon^2}v'_d ,\quad
M_D=\Mat3{0}{1}{1}{0}{1}{1}{0}{1}{1}\epsilon^2v_u , \quad
M_M=\Mat3{1}{0}{0}{0}{1}{1}{0}{1}{1}\frac{V^2}{M_P} .
}
Redefining the K\"ahler potential given by
\eqn{
K(E^c)&=&|E^c_1|^2+|E^c_2|^2+|E^c_3|^2
+\L\{
\frac{(E_2E^c_1)^*E^c_2}{M^2_P}
+\frac{(E_3E^c_1)^*E^c_3}{M^3_P}
+\frac{(V_2E^c_2)^*\cdot(V_1E^c_3)}{M^3_P}+h.c.\R\} \no \\
&=&((E^c_1)^*,(E^c_2)^*,(E^c_3)^*)
\Mat3{1}{\epsilon^2}{\epsilon^3}
{\epsilon^2}{1}{\epsilon^3}
{\epsilon^3}{\epsilon^3}{1}
\3tvec{E^c_1}{E^c_2}{E^c_3} , \\
K(L)&=&(|L_1|^2+|L_2|^2)+|L_3|^2 \no \\
&+&\L\{\frac{(L_1D_2+L_2D_1)^*(D_1L_3)+(L_1D_1-L_2D_2)^*(D_2L_3)}{M^2_P}
+\frac{|L\cdot V_1|^2}{M^2_P}+\cdots+h.c. \R\} \no \\
&=&(L^*_1,L^*_2,L^*_3)
\Mat3{1}{\epsilon^2}{\epsilon^2}
{\epsilon^2}{1}{\epsilon^2}
{\epsilon^2}{\epsilon^2}{1}
\3tvec{L_1}{L_2}{L_3} ,\\
K(N^c)&=&|N^c_1|^2+|N^c_2|^2+|N^c_3|^2
+\L\{
\frac{(V_1N^c_2)^*\cdot(V_1N^c_3)}{M^2_P}+h.c.\R\} \no \\
&=&((N^c_1)^*,(N^c_2)^*,(N^c_3)^*)
\Mat3{1}{0}{0}
{0}{1}{\epsilon^2}
{0}{\epsilon^2}{1}
\3tvec{N^c_1}{N^c_2}{N^c_3},
}
by the superfields redefinition as 
\eqn{
\3tvec{E^c_1}{E^c_2}{E^c_3}&=& V_K(E)
\3tvec{(E^c_1)'}{(E^c_2)'}{(E^c_3)'} ,\quad
V_K(E)=
\Mat3{1}{\epsilon^2}{\epsilon^3}
{\epsilon^2}{1}{\epsilon^3}
{\epsilon^3}{\epsilon^3}{1} ,\\
\3tvec{L_1}{L_2}{L_3}&=&V_K(L)
\3tvec{L'_1}{L'_2}{L'_3},\quad
V_K(L)=
\Mat3{1}{\epsilon^2}{\epsilon^2}
{\epsilon^2}{1}{\epsilon^2}
{\epsilon^2}{\epsilon^2}{1} ,\\
\3tvec{N^c_1}{N^c_2}{N^c_3}&=&V_K(N)
\3tvec{(N^c_1)'}{(N^c_2)'}{(N^c_3)'},\quad
V_K(N)=
\Mat3{1}{0}{0}
{0}{1}{\epsilon^2}
{0}{\epsilon^2}{1},
}
the modified mass matrices are given by
\eqn{
M'_e&=&V^T_K(L)
\Mat3{\epsilon^5}{\epsilon^3}{\epsilon^2}
{\epsilon^5}{\epsilon^3}{\epsilon^2}
{\epsilon^5}{\epsilon^3}{\epsilon^2}
v'_d
V_K(E)=
\Mat3{\epsilon^5}{\epsilon^3}{\epsilon^2}
{\epsilon^5}{\epsilon^3}{\epsilon^2}
{\epsilon^5}{\epsilon^3}{\epsilon^2}v'_d  , \\
M'_D&=&V^T_K(L)
\Mat3{0}{1}{1}{0}{1}{1}{0}{1}{1}\epsilon^3v_u
V_K(N)
=\Mat3{0}{1}{1}{0}{1}{1}{0}{1}{1}\epsilon^3v_u , \\
M'_M&=&V^T_K(N)
\Mat3{1}{0}{0}{0}{1}{1}{0}{1}{1}\frac{V^2}{M_P}
V_K(N)
=\Mat3{1}{0}{0}{0}{1}{1}{0}{1}{1}\frac{V^2}{M_P}.
}
The mixing matrices of charged leptons are given by 
\eqn{
M'_e&=&L_eM^{\mbox{diag}}_eR^\dagger_e
=\Mat3{1}{1}{1}{1}{1}{1}{1}{1}{1}
\Mat3{\epsilon^5}{0}{0}{0}{\epsilon^3}{0}{0}{0}{\epsilon^2}v'_d
\Mat3{1}{\epsilon^2}{\epsilon^3}
{\epsilon^2}{1}{\epsilon}
{\epsilon^3}{\epsilon}{1}.
}
The Yukawa hierarchy of charged leptons gives good agreement with
the experimental values given by
\eqn{
Y_e(M_P)&=&\frac{1}{1.9}\L(\frac{0.496\times 10^{-3}}{87}\R)=0.30\epsilon^5 , \\
Y_\mu(M_P)&=&\frac{1}{1.9}\L(\frac{105\times 10^{-3}}{87}\R)=0.64\epsilon^3 , \\
Y_\tau(M_P)&=&\frac{1}{1.9}\L(\frac{1.78}{87}\R)=1.08\epsilon^2 ,
}
where the used value of renormalization factor
\eqn{
\sqrt{\frac{\alpha_e(M_S)}{\alpha_e(M_P)}}=1.9
}
is calculated based on RGEs given in appendix A.

The neutrino seesaw mass matrix is given by
\eqn{
M_\nu=(M'_D)(M'_M)^{-1}(M'_D)^T=m_\nu\Mat3{1}{1}{1}{1}{1}{1}{1}{1}{1}, \quad
m_\nu=\frac{(\epsilon^3v_u)^2}{M_R}=O(0.01\mbox{eV}), \quad M_R=\frac{V^2}{M_P} ,
}
which has one zero eigenvalue because one RHN $n^c_1$ does not couple to left-handed leptons.
Therefore mixing matrix and mass eigenvalues are given as follows 
\eqn{
L^T_\nu M_\nu L_\nu&=&\mbox{diag}(m_{\nu_1},m_{\nu_2},m_{\nu_3}) , \\
L_\nu&=&\Mat3{1}{1}{1}{1}{1}{1}{1}{1}{1} \\
m_{\nu_1}=0,&& m_{\nu_2}=\sqrt{m^2_{21}}=0.87\times 10^{-2},\quad
m_{\nu_3}\simeq\sqrt{m^2_{32}}=4.8\times 10^{-2} \quad (\mbox{eV}),
}
and MNS matrix is given by
\eqn{
V_{MNS}=L^\dagger_eL_\nu=\Mat3{O(1)}{O(1)}{O(1)}{O(1)}{O(1)}{O(1)}{O(1)}{O(1)}{O(1)}.
}
With the recent experimental value of $|\sin\theta_{13}|\sim 0.14$ \cite{Tortola:2012te},
MNS matrix is given by
\eqn{
V_{MNS}=\Mat3{0.64}{0.55}{0.14}
{0.42}{0.64}{0.65}
{0.36}{0.55}{0.76},
}
which requires accidental cancellation of two mixing matrices $L_{e,\nu}$
to reproduce the small $\theta_{13}$.

\subsection{Squark and slepton sector}

Sfermion mass matrices are given as follows
\eqn{
-{\cal L}&\supset& \sum_{X=Q,U^c,D^c,L,E^c}X^*_a[M^2(X)]_{ab}X_b  \\
&-&[H^U_3Q_aA(U)_{ab}U^c_b+H^D_3Q_aA(D)_{ab}D^c_b
+H^D_3L_aA(E)_{ab}E^c_b+h.c.]
+V_F+V_D ,\no\\
M^2(Q)&=&
\Mat3{m^2_Q}{m^2\epsilon^2}{m^2\epsilon^2}
{m^2\epsilon^2}{m^2_Q}{m^2\epsilon^2}
{m^2\epsilon^2}{m^2\epsilon^2}{m^2_{Q_3}} ,\\
M^2(U^c)&=&
\Mat3{m^2_{U^c_1}}{m^2\epsilon^3}{m^2\epsilon^4}
{m^2\epsilon^3}{m^2_{U^c_2}}{m^2\epsilon^5}
{m^2\epsilon^4}{m^2\epsilon^5}{m^2_{U^c_3}} ,\\
M^2(D^c)&=&
\Mat3{m^2_{D^c_1}}{m^2\epsilon^3}{m^2\epsilon^3}
{m^2\epsilon^3}{m^2_{D^c_2}}{m^2\epsilon^6}
{m^2\epsilon^3}{m^2\epsilon^6}{m^2_{D^c_3}} ,\\
M^2(L)&=&
\Mat3{m^2_L}{m^2\epsilon^2}{m^2\epsilon^2}
{m^2\epsilon^2}{m^2_L}{m^2\epsilon^2}
{m^2\epsilon^2}{m^2\epsilon^2}{m^2_{L_3}} ,\\
M^2(E^c)&=&
\Mat3{m^2_{E^c_1}}{m^2\epsilon^2}{m^2\epsilon^3}
{m^2\epsilon^2}{m^2_{E^c_2}}{m^2\epsilon^3}
{m^2\epsilon^3}{m^2\epsilon^3}{m^2_{E^c_3}} ,\\
A(U)&=&
\Mat3{\epsilon^6Y^UA^U}{\epsilon^3Y^UA^U}{\epsilon^2Y^UA^U}
{\epsilon^6Y^UA^U}{\epsilon^3Y^UA^U}{\epsilon^2Y^UA^U}
{\epsilon^4Y^UA^U}{0}{Y^U_3A^U_3} ,\\
A(D)&=&
\Mat3{\epsilon^5Y^DA^D}{\epsilon^4Y^DA^D}{\epsilon^4Y^DA^D}
{\epsilon^5Y^DA^D}{\epsilon^4Y^DA^D}{\epsilon^4Y^DA^D}
{\epsilon^3Y^DA^D}{0}{\epsilon^2Y^DA^D} ,\\
A(E)&=&
\Mat3{\epsilon^5Y^EA^E}{\epsilon^3Y^EA^E}{\epsilon^2Y^EA^E}
{\epsilon^5Y^EA^E}{\epsilon^3Y^EA^E}{\epsilon^2Y^EA^E}
{\epsilon^5Y^EA^E}{\epsilon^3Y^EA^E}{\epsilon^2Y^EA^E} ,\\
V_F&=&|Y^U_3H^U_3Q_3|^2+|Y^U_3H^U_3U^c_3|^2+|Y^U_3Q_3U^c_3+\lambda_3S_3H^D_3|^2 ,\\
V_D&=&\frac12 g^2_x\L[5|S_3|^2
+\sum^3_{a=1}(|Q_a|^2+|U^c_a|^2+2|D^c_a|^2+2|L_a|^2+|E^c_a|^2)\R]^2,
}
where $m=O(10^3\mbox{GeV})$ and the contributions from F-terms are neglected except for
top-Yukawa contributions and the contributions from D-terms
are neglected except for the contributions from $S_3$.

After the redefinition of K\"ahler potential and the diagonalization of
Yukawa matrices, sfermion masses are given as follows
\eqn{
-{\cal L}&\supset&\sum^2_{i=1}\L(m^2_{U^c_i}+5g^2_x(v'_s)^2\R)|U^c_i|^2
+\sum^3_{a=1}\L(m^2_{D^c_a}+10g^2_x(v'_s)^2\R)|D^c_a|^2 
+\sum^2_{i=1}\L(m^2_Q+5g^2_x(v'_s)^2\R)|Q_i|^2 \no \\
&+&\L(m^2_{Q_3}+5g^2_x(v'_s)^2\R)|D_3|^2
+\sum^2_{i=1}\L(m^2_L+10g^2_x(v'_s)^2\R)|L_i|^2 \no \\
&+&\L(m^2_{L_3}+10g^2_x(v'_s)^2\R)|L_3|^2
+\sum^3_{a=1}\L(m^2_{E^c_a}+5g^2_x(v'_s)^2\R)|E^c_a|^2 \no \\
&+&(U^*_3,U^c_3)
\mat2{m^2_{Q_3}+(Y^U_3v'_u)^2+5g^2_x(v'_s)^2}{Y^U_3\lambda_3v'_sv'_d-A^U_3Y^U_3v'_u}
{Y^U_3\lambda_3v'_sv'_d-A^U_3Y^U_3v'_u}{m^2_{U_3}+(Y^U_3v'_u)^2+5g^2_x(v'_s)^2}
\2tvec{U_3}{(U^c_3)^*} \no \\
&+&m^2U^*_a(\delta^U_{LL})_{ab}U_b+m^2D^*_a(\delta^D_{LL})_{ab}D_b
+m^2(U^c)^*_a(\delta^U_{RR})_{ab}U^c_b
+m^2(D^c)^*_a(\delta^D_{RR})_{ab}D^c_b \no \\
&+&m^2E^*_a(\delta^E_{LL})_{ab}E_b+m^2N^*_a(\delta^N_{LL})_{ab}N_b
+m^2(E^c)^*_a(\delta^E_{RR})_{ab}E^c_b \no \\
&-&m^2\L\{U_a(\delta^U_{LR})_{ab}U^c_b
+D_a(\delta^D_{LR})_{ab}D^c_b
+E_a(\delta^E_{LR})_{ab}E^c_b+h.c. \R\} ,}
\eqn{
\delta^U_{LL}&=&\frac{1}{m^2}[L^\dagger_uV^\dagger_K(Q)M^2(Q)V_K(Q)L_u]_{\mbox{off diagonal}}
=\Mat3{0}{\epsilon^2}{\epsilon^2}{\epsilon^2}{0}{\epsilon^2}{\epsilon^2}{\epsilon^2}{0} , \\
\delta^D_{LL}&=&\frac{1}{m^2}[L^\dagger_dV^\dagger_K(Q)M^2(Q)V_K(Q)L_d]_{\mbox{off diagonal}}
=\Mat3{0}{\epsilon^2}{\epsilon^2}{\epsilon^2}{0}{\epsilon^2}{\epsilon^2}{\epsilon^2}{0} , \\
\delta^E_{LL}&=&\frac{1}{m^2}[L^\dagger_eV^\dagger_K(L)M^2(L)V_K(L)L_e]_{\mbox{off diagonal}}
=\Mat3{0}{1}{1}{1}{0}{1}{1}{1}{0} , \\
\delta^U_{RR}&=&\frac{1}{m^2}[R^\dagger_uV_K^\dagger(U)M^2(U)V_K(U)R_u]_{\mbox{off diagonal}}
=\Mat3{0}{\epsilon^2}{\epsilon^4}{\epsilon^2}{0}{\epsilon^5}{\epsilon^4}{\epsilon^5}{0} , \\
\delta^D_{RR}&=&\frac{1}{m^2}[R^\dagger_dV_K^\dagger(D)M^2(D)V_K(D)R_d]_{\mbox{off diagonal}}
=\Mat3{0}{\epsilon^3}{\epsilon}{\epsilon^3}{0}{\epsilon^4}{\epsilon}{\epsilon^4}{0} , \\
\delta^U_{LR}&=&\frac{1}{m^2}[L^T_uV^T_K(Q)A(U)V_K(U)R_u]_{A^U_3=0}
=\frac{v'_uY^UA^U}{m^2}
\Mat3{\epsilon^6}{\epsilon^3}{\epsilon^2}
{\epsilon^6}{\epsilon^3}{\epsilon^2}
{\epsilon^4}{\epsilon^5}{0} , \\
\delta^D_{LR}&=&\frac{1}{m^2}[L^T_dV^T_K(Q)A(D)V_K(D)R_d]
=\frac{v'_dY^DA^D}{m^2}
\Mat3{\epsilon^5}{\epsilon^4}{\epsilon^4}
{\epsilon^5}{\epsilon^4}{\epsilon^4}
{\epsilon^3}{\epsilon^6}{\epsilon^2} , \\
\delta^E_{LR}&=&\frac{1}{m^2}[L^T_eV^T_K(L)A(E)V_K(E)R_e]
=\frac{v'_dY^EA^E}{m^2}
\Mat3{\epsilon^5}{\epsilon^3}{\epsilon^2}
{\epsilon^5}{\epsilon^3}{\epsilon^2}
{\epsilon^5}{\epsilon^3}{\epsilon^2},
}
where the off diagonal parts are extracted except for stop mass matrix
and $\delta^N_{LL},\delta^E_{RR}$ are omitted.

\subsection{Flavor and CP violation}

The off diagonal elements of sfermion mass matrices
contribute to flavor and CP violation
through the sfermion exchange, on which are imposed severe constraints.
Based on the estimations of the flavor and CP violations
with the mass insertion approximation,
the upper bounds for each elements are given in Table 4,
where
$M_Q=M(\mbox{gluino})=M(\mbox{squark}),
M_L=M(\mbox{slepton})=M(\mbox{photino})$ are assumed \cite{SUSYFCNC}.
Note that there is another suppression factor in $\delta^X_{LR}$ as
\eqn{
\frac{v'_{u,d}}{m}\sim \epsilon .
}

The most stringent bound for $M_L$ is given by $\mu\to e\gamma$ as
\eqn{
1<1.5\times 10^{-2}\L(\frac{M_L}{300\mbox{GeV}}\R)^2\quad :\quad
M_L>2250\mbox{GeV},
}
and the one for $M_Q$ is given by $\epsilon_K$ as
\eqn{
\epsilon^{2.5}=3\times 10^{-3}<4.4\times 10^{-4}
\L(\frac{M_Q}{1000\mbox{GeV}}\R)\quad :\quad
M_Q>6820\mbox{GeV}.
}
Note that if $Q_{1,2}$ were $S_4$-singlets, then $(\delta^U_{LL})_{12}$ would be $O(1)$
and the most stringent bound for $M_Q$ would be given by
\eqn{
1<6.4\times 10^{-3}\L(\frac{M_Q}{1000\mbox{GeV}}\R)\quad :\quad
M_Q>156\mbox{TeV}.
}
Comparing Eq.(182) and Eq.(183), one can see that $S_4$ softens the SUSY flavor problem
very efficiently.

Before ending this section, we discuss the problem of a complex flavon VEV.
If the relative phase of two VEVs $\L<D_1\R>,\L<D_2\R>$ exists,
we must include
\eqn{
K(D^c)&\supset&\L\{\frac{[-D^*_1D_2+D^*_2D_1](D^c_2)^*(D^c_3)}{M^2_P}+h.c.\R\} ,
}
in K\"ahler potential, then redefinition of superfields are modified as follows
\eqn{
\3tvec{D^c_1}{D^c_2}{D^c_3}=K'(D)
\3tvec{(D^c_1)'}{(D^c_2)'}{(D^c_3)'},\quad
V_K(D)=
\Mat3{1}{\epsilon^3}{\epsilon^3}
{\epsilon^3}{1}{\epsilon^2}
{\epsilon^3}{\epsilon^2}{1}.
}
Therefore the mass matrix and mixing matrix of down quark sector 
and off-diagonal matrix of squarks are
modified as follows
\eqn{
M'_d=
\Mat3{\epsilon^5}{\epsilon^4}{\epsilon^4}
{\epsilon^5}{\epsilon^4}{\epsilon^4}
{\epsilon^3}{\epsilon^4}{\epsilon^2},\quad
R'_d=
\Mat3{1}{\epsilon}{\epsilon}
{\epsilon}{1}{\epsilon^2}
{\epsilon}{\epsilon^2}{1}, \quad
\delta^D_{RR}
=\Mat3{0}{\epsilon}{\epsilon}
{\epsilon}{0}{\epsilon^2}
{\epsilon}{\epsilon^2}{0}.
}
As the result, the most stringent bound for $M_Q$
is changed into $M_Q>68\mbox{TeV}$.
This suggests new mechanism is needed to suppress CP violation.
We leave this problem for future work.

\begin{table}[htbp]
\begin{center}
\begin{tabular}{|c|c|c|c|c|}
\hline
observable      &parameter   &order  &upper bound &coefficient \\ \hline
$\Delta m_K$    &
\begin{tabular}{c}
$\sqrt{(\delta^D_{LL})_{12}(\delta^D_{RR})_{12}}$ \\ 
$(\delta^D_{LL})_{12}$ \\ 
$(\delta^D_{LR})_{12}$ \\ 
\end{tabular}   &
\begin{tabular}{c}
$\epsilon^{2.5}$ \\
$\epsilon^2$ \\
$\epsilon^4$ \\
\end{tabular} &
\begin{tabular}{c}
$5.6\times 10^{-3}$ \\
$8.0\times 10^{-2}$  \\
$8.8\times 10^{-3}$ \\
\end{tabular}   &
$\times\L(\frac{M_Q}{1000GeV}\R)
\sqrt{\frac{(\Delta m_K)_{exp}}{3.49\times 10^{-12}MeV}}$ 
\\ \hline
$\Delta m_B$    &
\begin{tabular}{c}
$\sqrt{(\delta^D_{LL})_{13}(\delta^D_{RR})_{13}}$ \\ 
$(\delta^D_{LL})_{13}$ \\ 
$(\delta^D_{LR})_{13}$ \\ 
\end{tabular}   &
\begin{tabular}{c}
$\epsilon^{1.5}$ \\
$\epsilon^2$ \\
$\epsilon^3$ \\
\end{tabular} &
\begin{tabular}{c}
$3.4\times 10^{-2}$ \\
$1.9\times 10^{-1}$ \\
$6.3\times 10^{-2}$ \\
\end{tabular}   &
$\times\L(\frac{M_Q}{1000GeV}\R)
\sqrt{\frac{(\Delta m_B)_{exp}}{3.38\times 10^{-10}MeV}}$ 
\\ \hline
$\Delta m_D$    &
\begin{tabular}{c}
$\sqrt{(\delta^U_{LL})_{12}(\delta^U_{RR})_{12}}$ \\ 
$(\delta^U_{LL})_{12}$ \\ 
$(\delta^U_{LR})_{12}$ \\ 
\end{tabular}   &
\begin{tabular}{c}
$\epsilon^2$ \\
$\epsilon^2$ \\
$\epsilon^3$ \\
\end{tabular} &
\begin{tabular}{c}
$1.1\times 10^{-2}$ \\
$6.2\times 10^{-2}$ \\
$1.9\times 10^{-2}$ \\
\end{tabular}   &
$\times\L(\frac{M_Q}{1000GeV}\R)
\sqrt{\frac{(\Delta m_B)_{exp}}{1.26\times 10^{-11}MeV}}$ 
\\ \hline
$\epsilon_K$    &
\begin{tabular}{c}
$\sqrt{|\mbox{Im}[(\delta^D_{LL})_{12}(\delta^D_{RR})_{12}]|}$ \\
$\sqrt{|\mbox{Im}(\delta^D_{LL})^2_{12}|}$  \\
$\sqrt{|\mbox{Im}(\delta^D_{LR})^2_{12}|}$  \\
\end{tabular}   &
\begin{tabular}{c}
$\epsilon^{2.5}$ \\
$\epsilon^2$ \\
$\epsilon^4$ \\
\end{tabular} &
\begin{tabular}{c}
$4.4\times 10^{-4}$ \\
$6.4\times 10^{-3}$ \\
$7.0\times 10^{-4}$ \\
\end{tabular}   &
$\times\L(\frac{M_Q}{1000GeV}\R)
\sqrt{\frac{(\epsilon_K)_{exp}}{2.24\times 10^{-3}}}$
\\ \hline
$\mu\to e\gamma$&
\begin{tabular}{c}
$(\delta^E_{LL})_{12}$ \\
$(\delta^E_{LR})_{12}$ \\
\end{tabular}   &
\begin{tabular}{c}
$1$        \\
$\epsilon^3$ \\
\end{tabular} &
\begin{tabular}{c}
$1.5\times 10^{-2}$ \\
$3.4\times 10^{-6}$ \\
\end{tabular}   &
$\times\L(\frac{M_L}{300GeV}\R)^2
\sqrt{\frac{(BR(\mu\to e\gamma))_{exp}}{2.4\times 10^{-12}}}$
\\ \hline
$\tau\to e\gamma$&
\begin{tabular}{c}
$(\delta^E_{LL})_{13}$ \\
$(\delta^E_{LR})_{13}$ \\
\end{tabular}   &
\begin{tabular}{c}
$1$          \\
$\epsilon^2$ \\
\end{tabular} &
\begin{tabular}{c}
$4.3$                \\
$1.6\times 10^{-2}$ \\
\end{tabular}   &
$\times\L(\frac{M_L}{300GeV}\R)^2
\sqrt{\frac{(BR(\tau\to e\gamma))_{exp}}{3.3\times 10^{-8}}}$
\\ \hline
$\tau\to \mu\gamma$ &
\begin{tabular}{c}
$(\delta^E_{LL})_{23}$ \\
$(\delta^E_{LR})_{23}$ \\
\end{tabular}   &
\begin{tabular}{c}
$1$          \\
$\epsilon^2$ \\
\end{tabular} &
\begin{tabular}{c}
$4.9$ \\
$1.8\times 10^{-2}$ \\
\end{tabular}   &
$\times\L(\frac{M_L}{300GeV}\R)^2
\sqrt{\frac{(BR(\tau\to \mu\gamma))_{exp}}{4.4\times 10^{-8}}}$
\\ \hline
$d_n$           &
\begin{tabular}{c}
$|\mbox{Im}(\delta^U_{LR})_{11}|$ \\
$|\mbox{Im}(\delta^D_{LR})_{11}|$ \\
\end{tabular}   &
\begin{tabular}{c}
$\epsilon^6$ \\
$\epsilon^5$ \\
\end{tabular} &
\begin{tabular}{c}
$3.1\times 10^{-6}$ \\
$1.6\times 10^{-6}$ \\
\end{tabular}   &
$\times\L(\frac{M_Q}{1000GeV}\R)
\L(\frac{(d_n)_{exp}}{2.9\times 10^{-26}ecm}\R)$
\\ \hline
$d_e$&$|\mbox{Im}(\delta^E_{LR})_{11}|$&$\epsilon^5$ &$1.7\times 10^{-7}$&
$\times\L(\frac{M_L}{300GeV}\R)
\L(\frac{(d_e)_{exp}}{1.05\times 10^{-27}ecm}\R)$ \\ \hline
\end{tabular}
\end{center}
\caption{Experimental constraints for the off diagonal elements of sfermion mass matrices
from meson mass splittings $\Delta m_K,\Delta m_B,\Delta m_D$, CP violating parameter $\epsilon_K$, 
lepton flavor violations $l_i\to l_j\gamma$ and electric dipole moments of neutron  $d_n$
and electron $d_e$. The predictions of our model for each parameters are given
in "order" column. The dependences of each upper bounds on experimental values are given in
"coefficient" column.}
\end{table}



\section{Cosmological Aspects}

Based on our model, we consider the scenario to reproduce
the cosmological parameters given as follows \cite{PDG2012}
\eqn{
\Omega_0&\simeq&\Omega_\Lambda+\Omega_b+\Omega_{CDM}\simeq 1, \\
\Omega_\Lambda&=&0.73\pm 0.03 . \\
\Omega_bh^2&=&0.0225\pm 0.0006 , \\
\Omega_{CDM}h^2&=&0.112\pm 0.006 , \\
h&=&0.704\pm 0.025 . 
}
For $\Omega_b$, we adopt leptogenesis as the mechanism to generate baryon asymmetry. 
For $\Omega_{CDM}$, we assume that dark matter consists of singlino dominated neutralino.

\subsection{Leptogenesis}

In general, leptogenesis scenario to generate baryon asymmetry
causes over production of gravitino in supersymmetric model.
This problem can be avoided in the case neutrino mass is
generated by small VEV of neutrinophilic Higgs doublet  \cite{nhiggs}.

In the diagonal RHN mass basis, superpotential of RHN is given by
\eqn{
W_N&=&\sum_{i=1,2}\epsilon^3H^U_i(L_1,L_2,L_3)
\Mat3{0}{Y^N_{i,12}}{Y^N_{i,13}}
{0}{Y^N_{i,22}}{Y^N_{i,23}}
{0}{Y^N_{i,32}}{Y^N_{i,33}}
\3tvec{N^c_1}{N^c_2}{N^c_3}
+\frac12 \sum^3_{a=1}M_aN^c_aN^c_a,
}
where we assume accidental mass hierarchy as follows
\eqn{
M_1=10^{3.5},\quad
M_2=M_3=10^4 \quad (\mbox{GeV}).
}
Note that these particles are enough light to create in low reheating temperature such
as $10^7\mbox{GeV}$ without causing gravitino over production \cite{reheating}. 
The interactions of right-handed sneutrinos (RHsNs) are given by
\eqn{
-{\cal L}_N&=&\sum_{i=1,2}\epsilon^3H^U_i(L_1,L_2,L_3)
\Mat3{0}{Y^N_{i,12}M_2}{Y^N_{i,13}M_3}
{0}{Y^N_{i,22}M_2}{Y^N_{i,23}M_3}
{0}{Y^N_{i,32}M_2}{Y^N_{i,33}M_3}
\3tvec{N^c_1}{N^c_2}{N^c_3} \no \\
&+&\sum_{i=1,2}\epsilon^3h^U_i(l_1,l_2,l_3)
\Mat3{0}{Y^N_{i,12}}{Y^N_{i,13}}
{0}{Y^N_{i,22}}{Y^N_{i,23}}
{0}{Y^N_{i,32}}{Y^N_{i,33}}
\3tvec{N^c_1}{N^c_2}{N^c_3},
}
where the contributions from A-terms are neglected. 
The $Z^N_2$ breaking scalar squared mass terms
\eqn{
K&\supset&\frac{F^*_{B_+}F_{B_{2-}}}{M^2_P}[(N^c_1)^*N^c_2+\cdots]+h.c. 
=\epsilon m^2[(N^c_1)^*N^c_2+\cdots]+h.c
}
fill in the zeros of sneurino mass matrix and gives
\eqn{
\Mat3{M^2_1}{\epsilon m^2}{\epsilon m^2}
{\epsilon m^2}{M^2_2}{m^2}
{\epsilon m^2}{m^2}{M^2_3}\sim 
M^2_2\Mat3{\epsilon}{\epsilon^3}{\epsilon^3}
{\epsilon^3}{1}{\epsilon^2}
{\epsilon^3}{\epsilon^2}{1},
}
where $O(\epsilon)$ suppressions of $Z^N_2$ breaking terms are assumed
without any reason. Note that the $O(\epsilon^3)$ elements are
originated from small $Z^N_2$ breaking parameters and small $Y^\Phi$ as 
\eqn{
\frac{m^2\epsilon}{m^2}\L(\frac{m}{M_2}\R)^2\sim \epsilon(Y^\Phi)^2
\sim \epsilon^3.
} 

In the diagonal RHsN mass basis, the interaction terms given in Eq.(194) are rewritten by
\eqn{
-{\cal L}_N&=&\sum_{i=1,2}M_2\epsilon^3H^U_i(L_1,L_2,L_3)
\Mat3{\epsilon^3Y^N_{i,11}}{Y^N_{i,12}}{Y^N_{i,13}}
{\epsilon^3Y^N_{i,21}}{Y^N_{i,22}}{Y^N_{i,23}}
{\epsilon^3Y^N_{i,31}}{Y^N_{i,32}}{Y^N_{i,33}}
\3tvec{N^c_1}{N^c_2}{N^c_3} \no \\
&+&\sum_{i=1,2}\epsilon^3h^U_i(l_1,l_2,l_3)
\Mat3{\epsilon^3Y^N_{i,11}}{Y^N_{i,12}}{Y^N_{i,13}}
{\epsilon^3Y^N_{i,21}}{Y^N_{i,22}}{Y^N_{i,23}}
{\epsilon^3Y^N_{i,31}}{Y^N_{i,32}}{Y^N_{i,33}}
\3tvec{N^c_1}{N^c_2}{N^c_3}.
}
The lightest RHN $n^c_1$ does not receive above corrections and remains decoupled.
Therefore lepton asymmetry is generated by the out of equilibrium decay of
the lightest RHsN $N^c_1$.

Following \cite{leptogenesis}, the CP asymmetry of 
sneutrino $N^c_1$ decay is calculated as follows
\eqn{
\epsilon_1&=&-\frac{1}{4\pi}\sum_k\frac{\mbox{Im}[K^2_{1k}]}{K_{11}}g(x_k) , \\
g(x)&=&\sqrt{x}\ln\frac{1+x}{x}+\frac{2\sqrt{x}}{x-1} , \\
x_k&=&\frac{M^2_k}{M^2_1} , \\
K_{ij}&=&\sum_{h=1,2}\sum^3_{l=1}(Y^N_{h,li})(Y^N_{h,lj})^* .
}
From the naive power counting, we obtain
\eqn{
K_{11}\sim \epsilon^{12},\quad K_{12}\sim K_{13}\sim \epsilon^9, \quad
\epsilon_1\sim \epsilon^6 .
}
Using $\epsilon_1$, the $B-L$ asymmetry generated via thermal leptogenesis is
expressed as
\eqn{
-(B-L)_f&=&\kappa\frac{\epsilon_1}{g_*} , \quad  g_*=341.25,
}
where $g_*$ is the total number of relativistic degrees of freedom contributing to the
energy density of the universe
and dilution factor $\kappa$ is defined as follows
\eqn{
\kappa&\sim& \frac{O(0.1)}{K} , \\
K&=&\frac{\Gamma(M_1)}{2H(M_1)} ,\\
\Gamma(M_1)&=&\frac{K_{11}M_1}{8\pi} ,\\
H(M_1)&=&\sqrt{\frac{\pi^2 g_*M^4_1}{90M^2_P}}.
}
By the EW sphaleron processes, the $B-L$ asymmetry is transferred to a 
B asymmetry as
\eqn{
B_f=\frac{24+4N_H}{66+13N_H}(B-L)_f\sim \frac13(B-L)_f,
}
where $N_H$ is number of Higgs doublets which are in 
equilibrium through Yukawa interactions, for example
$N_H=1$ for SM and $N_H=2$ for MSSM.
In any way $N_H$-dependence is not important for our rough estimation.
For our parameter values, we obtain $K \sim O(1)$ and
\eqn{
B_f\sim 10^{-10},
}
which is consistent with observed value
\eqn{
\eta_B=7.04B_f=6.1\times 10^{-10}.
}
Requiring the effective interaction
\eqn{
{\cal L}_{\mbox{eff}}=\epsilon^6\frac{(H^U_iL_j)^2}{2M_2}
}
is decoupled in order to avoid too strong wash out, we impose the condition as follows
\eqn{
\Gamma\sim \frac{\epsilon^{12}T^3}{8\pi^3 M^2_2}< H=\sqrt{\frac{\pi^2 g_*T^4}{90M^2_P}} ,
}
which gives upper bound for temperature as
\eqn{
T< 10^4\mbox{GeV}.
}
This condition is always satisfied after the decay of $N^c_1$ starts.

This scenario is different from conventional one in the point that neutrino mass
\eqn{
m_\nu\sim \frac{10^{-6}v^2_u}{M_2}\sim
 \L(\frac{v_u}{\mbox{GeV}}\R)^2 0.1\mbox{eV} \sim O(0.01\mbox{eV}),
}
is realized by the small VEV $v_u=O(1\mbox{GeV})$.

\subsection{Dark matter}

Here we calculate the relic abundance of LSP
which corresponds to singlino dominated neutralino in our model \cite{u1dm}.
The most dominant contribution to annihilation cross section of LSP
is given by the interaction with Z boson.
If the mass matrix given in Eq.(97) is diagonalized by the field redefinition as
\eqn{
\3tvec{(h^U_i)^0}{(h^D_i)^0}{s_i}
=\Mat3{V_a}{*}{*}
{V_b}{*}{*}
{V_c}{*}{*}
\3tvec{\chi^0_{i,1}}{\chi^0_{i,2}}{\chi^0_{i,3}},\quad
m_{\chi^0_{i,1}}<m_{\chi^0_{i,2}}<m_{\chi^0_{i,3}}, \quad (i=1,2),
}
the interaction with Z boson is given by
\eqn{
{\cal L}_Z&=&G(\chi^0_{1,1})\bar{\chi}^0_{i,1}Z_\mu\bar{\sigma}^\mu\chi^0_{i,1}
+iG(f_L)\bar{f}\gamma^\mu Z_\mu P_Lf
+iG(f_R)\bar{f}\gamma^\mu Z_\mu P_Rf, \\
&&G(\chi^0_{1,1})=\frac12(V^2_a-V^2_b)\sqrt{g^2_Y+g^2_2}=0.372(|V_a|^2-|V_b|^2) \\
&&G(e_L)=0.200,\quad G(e_R)=-0.172,\quad
G(\nu_L)=-0.372 ,\no \\
&&G(u_L)=-0.257 ,\quad G(u_R)=0.115 ,\quad
G(d_L)=0.314,\quad G(d_R)=-0.057,
}
where
\eqn{
\alpha_Y(m_Z)=0.0101687, \quad \alpha_2(m_Z)=0.0338098
}
are used.

The formula for the relic abundance of cold dark matter is given by
\eqn{
\Omega_{CDM}h^2&=&\frac{8.76\times 10^{-11}g^{-\frac12}_*x_F}{(a+3b/x_F)\mbox{GeV}^2}  , \\
x_F&=&\ln\frac{0.0955m_Pm_{\chi^0_1}(a+6b/x_F)}{(g_*x_F)^\frac12} , \\
&&m_P=1.22\times 10^{19}\mbox{GeV} \no  , \\
&&g_*=72.25\quad (T_F=m_{\chi^0_1}/x_F<m_\tau) , \\
a&=&\sum_f\frac{c_f}{2\pi}G^2(\chi^0_{1,1})\L[\frac{m_f}{4m^2_{\chi^0_1}-m^2_Z}
[G(f_L)-G(f_R)]\R]^2 \\
b&=&\sum_f \frac{c_f}{3\pi}G^2(\chi^0_{1,1})
\L(\frac{m_{\chi^0_1}}{4m^2_{\chi^0_{1,1}}-m^2_Z}\R)^2
[G^2(f_L)+G^2(f_R)]-\frac34 a
}
where $m_f\ll m_{\chi^0}$ is assumed. Substituting the values given 
in Eq.(218) and Eq.(219) and following values
\eqn{
m_Z=91.1876, \quad
m_b=4.18 , \quad
m_\tau=1.777 \quad (\mbox{GeV}),\quad
r=\frac{2m_{\chi^0_{1,1}}}{m_Z}
}
in Eq.(224) and Eq.(225), we get
\eqn{
a&=&1.77\times 10^{-8}\L(\frac{G(\chi^0_{1,1})}{r^2-1}\R)^2 (\mbox{GeV}^{-2}) ,\\
b&=&6.436\times 10^{-6}\L(\frac{G(\chi^0_{1,1})r}{r^2-1}\R)^2
-0.013\times 10^{-6}\L(\frac{G(\chi^0_{1,1})}{r^2-1}\R)^2 (\mbox{GeV}^{-2}) ,
}
from which the formula is rewritten as follows
\eqn{
x_F&=&\ln\L[\L(\frac{G(\chi^0_{1,1})}{0.01}\R)^2
\L(0.0177+\frac{6}{x_F}(6.436r^2-0.013)\R)\frac{6.25\times 10^8 r}{(r^2-1)^2}\R]
-\frac12\ln (x_F) , \\
\Omega_{CDM}h^2&=&\frac{0.10306x_F}{\L(\frac{G(\chi^0_{1,1})}{0.01}\R)^2
\L(0.0177+\frac{3}{x_F}(6.436r^2-0.013)\R)\frac{1}{(r^2-1)^2}} ,
}
where quark and lepton masses are neglected except for bottom and $\tau$.
Since the two LSPs $\chi^0_{1,1},\chi^0_{2,1}$ have the same mass and the same interactions,
they have the same relic abundance. Therefore the required relic abundance of one LSP is
$\Omega_{CDM}h^2=0.055$.
For the allowed range given in Eq.(104), the required values for $\lambda_{4,5}$
to reproduce observed relic abundance of dark matter are given in Table 5.
The allowed ranges for $\lambda_{4,5}$ are very small.
Note that we should not impose LEP bound ($m_{\chi^0_1}>46\mbox{GeV}$) 
on this LSP, because
$Z\to\chi^0_{i,1}\chi^0_{i,1}$ is strongly suppressed by the factor
$(|V_a|^2-|V_b|^2)^2/2\sim 0.005$ and the contribution to invisible decay width
is negligible as follows  \cite{PDG2012}
\eqn{
&&\Gamma(Z\to\chi^0_{i,1}\chi^0_{i,1})
\sim (0.6\times 2/3)0.005\Gamma(\mbox{invisible})
\sim 1.0\mbox{MeV}, \\
&&\Gamma(\mbox{invisible})=499.0\pm 1.5 \mbox{MeV},
}
where phase space suppression factor $\sim 0.6$ and the ratio of
LSP number and neutrino number $2/3$ are multiplied.

\begin{table}[htbp]
\begin{center}
\begin{tabular}{|c|c|c|c|c|c|c|c|c|c|c|c|}
\hline
$\lambda_4$&$\lambda_5$&$m_{\chi^\pm_1}$ &$m_{\chi^0_{1,1}}$&$m_{\chi^0_{1,2}}$&$m_{\chi^0_{1,3}}$
&$V_a$   &$V_b$  &$x_F$ &$T_F$ &$\Omega_{CDM}h^2$\\ \hline
0.44&0.57&141.2    &36.47 & 142.38 &178.84 &0.3667 &0.2225 &22.90 &1.592 &0.0552    \\ \hline
0.42&0.56&130.1    &36.52 & 131.09 &167.61 &0.3721 &0.2321 &22.90 &1.595 &0.0550    \\ \hline
0.40&0.55&119.1    &36.61 & 119.91 &156.52 &0.3776 &0.2429 &22.90 &1.599 &0.0551    \\ \hline
0.38&0.54&108.0    &36.79 & 108.63 &145.42 &0.3836 &0.2554 &22.91 &1.606 &0.0549    \\ \hline
0.37&0.53&102.5    &36.60 & 103.10 &139.69 &0.3885 &0.2591 &22.91 &1.597 &0.0550    \\ \hline
\end{tabular}
\end{center}
\caption{The parameter sets $(\lambda_4,\lambda_5,m^\pm_{\chi_1}=\lambda_1v'_s)$ 
which reproduce observed relic abundance of dark matter.
The dimensionful values are expressed in GeV units.}
\end{table}

\subsection{Constraint for long-lived  massive particles}

Finally we consider long-lived massive particles which are included in our model,
G-Higgs, flavons and the lightest RHN. Such particles are imposed on 
strong constraints from cosmological observations.

The superpotential of G-Higgs sector gives
degenerated G-higgsino mass as
\eqn{
M_g=kv'_s\mbox{diag}(1,1,1),
}
which receives $S_4$ breaking perturbation from 
K\"ahler potential given by
\eqn{
K(G)&=&|G_a|^2 +\frac{1}{M^2_P}\L[|2D_2G_1|^2+|(-\sqrt{3}D_1-D_2)G_2|^2
+|(\sqrt{3}D_1-D_2)G_3|^2\R]+(G\to G^c) \no \\
&=&|G_a|^2+\sum_ac_a\epsilon^2|G_a|^2+(G\to G^c),
}
which solves the mass degeneracy, however generation mixing is not induced.
Neglecting $O(\epsilon^2)$ corrections and contributions from D-terms
except for the contribution from $S_3$, the G-Higgs mass terms are given by
\eqn{
-{\cal L}&\supset&m^2_G|G_a|^2+m^2_{G^c}|G^c_a|^2
-\L[kA_kS_3G_aG^c_a+h.c.\R] +\L|kS_3G_a\R|^2+\L|kS_3G^c_a\R|^2\no \\
&+&\L|kG_aG^c_a+\lambda_3H^U_3H^D_3\R|^2 
+\frac12 g^2_x\L[5|S_3|^2-2|G_a|^2-3|G^c_a|^2\R]^2,
}
from which we obtain three same $2\times 2$ matrices as
\eqn{
M^2_a(G)&=&\mat2{m^2_G+(kv'_s)^2-10g^2_x(v'_s)^2}{\lambda_3kv'_uv'_d-kA_kv'_s}
{\lambda_3kv'_uv'_d-kA_kv'_s}{m^2_{G^c}+(kv'_s)^2-15g^2_x(v'_s)^2}.
}
The mass spectrum of G-Higgs and G-higgsino is given in Table 7 and
the lightest particle of them is lighter G-Higgs scalar $G_-$.
The dominant contributions to the $G_-$ decay are given by
the superpotential
\eqn{
W&\supset&\frac{1}{M^2_P}Q_3Q_3\Phi^c_3\sum_a\Phi_aG_a
+\frac{1}{M^2_P}U^c_3E^c_3\Phi^c_3\sum_a\Phi_aG_a \no \\
&=&\frac{1}{\sqrt{3}}Y^{QQ}Q_3Q_3(G_1+G_2+G_3)
+\frac{1}{\sqrt{3}}Y^{UE}U^c_3E^c_3(G_1+G_2+G_3) , \\
&&Y^{QQ}= Y^{UE}= \L(\frac{\L<\Phi_3\R>}{M_P}\R)^2\sim 2\times 10^{-14} ,
}
from which we obtain
\eqn{
{\cal L}_G&=&\frac{1}{\sqrt{3}}A^{UE}_{RF}Y^{UE}(e^c_3u^c_3+u^c_3e^c_3)G_1
+\frac{1}{\sqrt{3}}A^{QQ}_{RF}Y^{QQ}(2u_3d_3+2d_3u_3)G_1.
}
For simplicity, we assume $G\sim G_-$ then decay width of $G_-$ is given by
\eqn{
\Gamma(G_-)=\frac{M(G_-)}{16\pi}\L[
2\L(\frac{1}{\sqrt{3}}A^{UE}_{RF}\R)^2
+4\L(\frac{2}{\sqrt{3}}A^{QQ}_{RF}\R)^2
\R](Y^{QQ})^2,
}
where the renormalization factors are calculated based on the RGEs given in appendix A
as follows
\eqn{
A^{UE}_{RF}=\sqrt{\frac{\alpha_{UE}(M_S)}{\alpha_{UE}(M_P)}}=4.9,\quad 
A^{QQ}_{RF}=\sqrt{\frac{\alpha_{QQ}(M_S)}{\alpha_{QQ}(M_P)}}=12.8.
}
Substituting these values in Eq.(240) we obtain the life time of $G_-$ as
\eqn{
\tau(G_-)&=&\frac{1}{\Gamma(G_-)}
=3.8\times 10^{-29}\L(\frac{M(G_-)}{1\mbox{TeV}}\R)^{-1}(Y^{QQ})^{-2}\quad \mbox{sec}.
}
Since the existence of a particle which has longer life time than 0.1 second
spoils the success of BBN \cite{reheating}, we must require $\tau(G_-)<0.1\mbox{sec}$
which impose constraint as
\eqn{
M(G_-)> \L(\frac{1.9\times 10^{-14}}{Y^{QQ}}\R)^2  \mbox{TeV}.
}
The G-Higgs exchange may contribute to proton decay, however it seems that the suppression
of power of $\epsilon$ is too strong to observe proton decay \cite{s4u1proton}.

The five of six flavon multiplets $\Phi_a,\Phi^c_a$ have $O(1\mbox{TeV})$ masses
which are enough small to product them non-thermally 
through the $U(1)_Z$ gauge interaction.
The lightest flavon (LF) is quasi-stable and should not produced so much
in order not to dominate $\Omega_{CDM}$. 
Solving the Boltzmann equation with the boundary condition $n_{LF}(T_{RH})=0$, 
we get relic abundance of LF as \cite{u1pamela} 
\footnote{Since the $U(1)_Z$ charge of $\Phi$ in \cite {u1pamela} is two times larger than
one in the present model, we multiply the equation for $\Omega_{LF}h^2$ 
given in \cite {u1pamela} by the factor $2^2$.}
\eqn{
\Omega_{LF}h^2=2.0\times 10^{-8}\L(\frac{T_{RH}}{10^5\mbox{GeV}}\R)^3
\L(\frac{10^{12}\mbox{GeV}}{\L<\Phi^c_3\R>}\R)^4
=2.0\times 10^{-6}\L(\frac{T_{RH}}{10^5\mbox{GeV}}\R)^3.
}
Requiring the LF does not dominate dark matter as $\Omega_{LF}h^2  < 0.01$, the upper bound for
reheating temperature is given by
\eqn{
T_R<10^6\mbox{GeV},
}
which is consistent with our leptogenesis scenario.

The life time of LF is estimated as follows.
The LF can decay, for example  through the operator
\eqn{
W=\frac{M_P(\epsilon^3H^U_iL_j)^2}{2(V+\Phi)^2}
\sim \frac{M_P(\epsilon^3H^U_iL_j)^2}{2V^2}\L(1-2\frac{\Phi}{V}\R),
}
the decay width and life time are given by \cite{u1pamela}
\eqn{
\Gamma(LF\to llHH)&=&\frac{M_S}{16\pi}\L(\frac{M^2_S(\epsilon^3)^2}{32\pi^2M_2V}\R)^2O(0.1)
\sim 10^{-29}\mbox{eV} , \\
\tau(LF)&\sim&10^{14}\mbox{sec}\sim 10^7\mbox{years},
}
which suggests LF does not exist in present universe. Note that three and two body decays
are suppressed by small VEV $v_u$.

The lightest RHN $n^c_1$ behaves like LF because there is no distinction between
$N^c$ and $\Phi^c$ under the gauge symmetry. Integrating out $N^c_1$ and $\lambda_Z$ 
in the Lagrangian
\eqn{
{\cal L}\supset g_Z\L(n^c_1\lambda_Z(N^c_1)^*+\psi\lambda_Z\Psi^*\R)
+\epsilon^6N^c_1lh^U_i ,
}
where $(\psi,\Psi)$ means super-multiplet and some factors are omitted for simplicity, we get
\eqn{
{\cal L}_{\mbox{eff}} =\frac{g^2_Z(\epsilon^6)}{(g_ZV)M^2_1}(n^c_1\psi)(lh^U_i)\Psi,
}
from which the life time of $n^c_1$ is given by
\eqn{
\Gamma(n^c_1\to \psi\Psi l h^U)&=&\frac{M^7_1}{16\pi(32\pi^2)^2}\L(\frac{g^2_Z(\epsilon^6)}{(g_ZV)M^2_1}\R)^2
\sim \L(\frac{M_1}{M_S}\R)^5\Gamma(LF\to llHH)
\sim 10^{-22}\mbox{eV} ,\\
\tau(n^c_1)&\sim& \L(\frac{M_S}{M_1}\R)^5\tau(LF)\sim
10^{12}\mbox{sec}\sim 10^5\mbox{years}.
}



\section{Conclusion}

In this paper we consider $S_4$ flavor symmetric extra U(1) model and obtain following results. 
\begin{itemize}
\item With the assignment of flavor representation to reproduce quark and lepton mass hierarchies
and mixing matrices, SUSY flavor problem is softened. 
\item Proton decay through G-Higgs exchange is suppressed by flavor symmetry.
\item Observed Higgs mass $125-126\mbox{GeV}$ is realized with stop lighter than $2\mbox{TeV}$
which is within  the testable range in LHC at $\sqrt{s}=14\mbox{TeV}$.
\item The partial gauge coupling unification at $M_P$ is realized by adding 4-th generation
Higgs and left-handed lepton which play the role to break $U(1)_Z$ gauge symmetry.
\item The allowed region for lightest chargino mass is given by $100-140 \mbox{GeV}$ when we
assume LSP is lightest singlino dominated neutralino. 
\item The extra Higgs doublets play the role of neutrinophilic Higgs which is needed for
low temperature leptogengesis without causing gravitino over production.
\item The shorter life time than 0.1 second of G-Higgs is realized. 
\item The over productions of flavon and lightest RHN are also avoided.
\end{itemize}

\section*{Acknowledgments}
H.O. thanks to Dr. Yuji Kajiyama and Dr. Kei Yagyu for fruitful discussion.


\appendix

\section{RGEs}

$O(1)$ coupling constants of our model consist of
gauge coupling constants and trilinear coupling constants defined by
\eqn{
W&\supset&\lambda_3S_3H^U_3H^D_3 
+\lambda_4H^U_3(S_1H^D_1+S_2H^D_2)+\lambda_5(S_1H^U_1+S_2H^U_2)H^D_3 \no \\
&+&kS_3(G_1G^c_1+G_2G^c_2+G_3G^c_3)+Y^U_3H^U_3Q_3U^c_3,
}
from which the fine structure constants are defined as follows
\eqn{
&&\alpha_Y=\frac{g^2_Y}{4\pi},\quad
\alpha_2=\frac{g^2_2}{4\pi},\quad
\alpha_3=\frac{g^2_3}{4\pi},\quad
\alpha_X=\frac{g^2_X}{4\pi},\quad
\alpha_Z=\frac{g^2_Z}{4\pi},\no \\
&&\alpha_t=\frac{(Y^U_3)^2}{4\pi},\quad
\alpha_h=\frac{\lambda^2_3}{4\pi},\quad
\alpha_4=\frac{\lambda^2_4}{4\pi},\quad
\alpha_5=\frac{\lambda^2_5}{4\pi},\quad
\alpha_k=\frac{k^2}{4\pi}.
}
We define the step functions as follows
\eqn{
&&\theta(x)=\L\{
\begin{array}{cc}
1& x\geq 0 \\
0& x<0 \\
\end{array} , 
\R. \\
&&\theta_I=\theta(\mu-M_I),\quad
\theta_4=\theta(\mu-M_{L_4}),\quad
\theta_5=\theta(\mu-M_{L_5}),\quad \\
&&M_I=10^{11.5}\mbox{GeV},\quad M_{L_4}=2.2\times 10^{14}\mbox{GeV}, \quad
M_{L_5}=2.4\times 10^{17}\mbox{GeV}. \no
}
The beta functions are given by
\eqn{
(2\pi)\frac{d\alpha_Y}{dt}
&=&\alpha^2_Y\L[15+20\frac{\alpha_3}{2\pi}+\frac{15}{2}\frac{\alpha_2}{2\pi}
+\L(2+3\frac{\alpha_2}{2\pi}\R)\theta_4+ \L(\frac{10}{3}+3\frac{\alpha_2}{2\pi}
+\frac{32}{9}\frac{\alpha_3}{2\pi}\R)\theta_5\R] , \\
(2\pi)\frac{d\alpha_2}{dt}
&=&\alpha^2_2\L[3+12\frac{\alpha_3}{2\pi}+\frac{39}{2}\frac{\alpha_2}{2\pi}
+\L(2+7\frac{\alpha_2}{2\pi}\R)\theta_4
+\L(2+7\frac{\alpha_2}{2\pi}\R)\theta_5\R]  , \\
(2\pi)\frac{d\alpha_3}{dt}
&=&\alpha^2_3\L[24\frac{\alpha_3}{2\pi}+\frac{9}{2}\frac{\alpha_2}{2\pi}
+\L(2+\frac{34}{3}\frac{\alpha_3}{2\pi}\R)\theta_5\R] ,  \\
(2\pi)\frac{d\alpha_X}{dt}
&=&\alpha^2_X\L[15+20\frac{\alpha_3}{2\pi}+\frac{15}{2}\frac{\alpha_2}{2\pi}
+\L(\frac{4}{3}+2\frac{\alpha_2}{2\pi}\R)\theta_4
+\L(\frac{10}{3}+2\frac{\alpha_2}{2\pi}
+\frac{16}{3}\frac{\alpha_3}{2\pi}\R)\theta_5\R]  , \\
(2\pi)\frac{d\alpha_Z}{dt}&=&\alpha^2_Z\L[\frac{65}{3}
+20\frac{\alpha_3}{2\pi}+\frac{15}{2}\frac{\alpha_2}{2\pi}
+\L(\frac{20}{9}+\frac{10}{3}\frac{\alpha_2}{2\pi}\R)\theta_4
+\L(\frac{50}{9}+\frac{10}{3}\frac{\alpha_2}{2\pi}
+\frac{80}{9}\frac{\alpha_3}{2\pi}\R)\theta_5\R]\theta_I, \\
t&=&\ln\mu,
}
where we include only the contributions from $\alpha_{2,3}$
in 2-loop order terms. 
The RGEs for trilinear coupling constants are given by
\eqn{
(2\pi)\frac{d\alpha_t}{dt}&=&\alpha_t\L(6\alpha_t+\alpha_h+2\alpha_4
-\frac{16}{3}\alpha_3-3\alpha_2-\frac{13}{9}\alpha_Y
-\frac12\alpha_X-\frac{5}{6}\alpha_Z\theta_I\R)  , \\
(2\pi)\frac{d\alpha_h}{dt}&=&\alpha_h
\L(3\alpha_t+4\alpha_h+2\alpha_4+2\alpha_5+9\alpha_k
-3\alpha_2-\alpha_Y
-\frac{19}{6}\alpha_X-\frac{5}{6}\alpha_Z\theta_I\R) , \\
(2\pi)\frac{d\alpha_4}{dt}&=&\alpha_4
\L(3\alpha_t+\alpha_h+5\alpha_4+2\alpha_5
-3\alpha_2-\alpha_Y
-\frac{19}{6}\alpha_X-\frac{5}{6}\alpha_Z\theta_I\R) , \\
(2\pi)\frac{d\alpha_5}{dt}&=&\alpha_5
\L(\alpha_h+2\alpha_4+5\alpha_5
-3\alpha_2-\alpha_Y
-\frac{19}{6}\alpha_X-\frac{5}{6}\alpha_Z\theta_I\R) , \\
(2\pi)\frac{d\alpha_k}{dt}&=&\alpha_k
\L(2\alpha_h+11\alpha_k
-\frac{16}{3}\alpha_3-\frac49\alpha_Y
-\frac{19}{6}\alpha_X-\frac{5}{6}\alpha_Z\theta_I\R) .
}
We define gaugino mass parameters and A-parameters as follows
\eqn{
{\cal L}&\supset &-\frac12 M_Y\lambda_Y\lambda_Y
-\frac12 M_2\lambda_2\lambda_2-\frac12 M_3\lambda^g_3\lambda^g_3
-\frac12 M_X\lambda_X\lambda_X-\frac12 M_Z\lambda_Z\lambda_Z \no \\
&+&\lambda_3A_3S_3H^U_3H^D_3 
+\lambda_4A_4H^U_3(S_1H^D_1+S_2H^D_2)+\lambda_5A_5(S_1H^U_1+S_2H^U_2)H^D_3 \no \\
&+&kA_kS_3(G_1G^c_1+G_2G^c_2+G_3G^c_3)+Y^U_3A_tH^U_3Q_3U^c_3  +h.c. .
}
The RGEs for gaugino mass parameters are given by
\eqn{
(2\pi)\frac{dM_Y}{dt}&=&\alpha_YM_Y\L[15+2\theta_4
+\frac{10}{3}\theta_5\R]  , \\
(2\pi)\frac{dM_2}{dt}&=&\alpha_2M_2\L[3+2\theta_4+2\theta_5\R] , \\
(2\pi)\frac{dM_3}{dt}&=&\alpha_3M_3\L[2\theta_5
+\frac{48}{2\pi}\alpha_3\R]  , \\
(2\pi)\frac{dM_X}{dt}&=&\alpha_XM_X\L[15+\frac43\theta_4
+\frac{10}{3}\theta_5\R]  , \\
(2\pi)\frac{dM_Z}{dt}&=&\alpha_ZM_Z\L[\frac{65}{3}
+\frac{20}{9}\theta_4+\frac{50}{9}\theta_5\R]\theta_I ,
}
where we take account of 2-loop contributions only for $M_3$.
The RGEs for A-parameters are given by
\eqn{
(2\pi)\frac{dA_t}{dt}&=&6\alpha_t A_t+\alpha_h A_3
+2\alpha_4A_4
+\frac{16}{3}\alpha_3M_3+3\alpha_2M_2+\frac{13}{9}\alpha_YM_Y \no \\
&+&\frac12\alpha_XM_X
+\frac{5}{6}\alpha_ZM_Z\theta_I , \\
(2\pi)\frac{dA_3}{dt}&=&3\alpha_t A_t+4\alpha_h A_3
+2\alpha_4A_4+2\alpha_5A_5+9\alpha_kA_k \no \\
&+&3\alpha_2M_2+\alpha_YM_Y+\frac{19}{6}\alpha_XM_X
+\frac{5}{6}\alpha_ZM_Z\theta_I , \\
(2\pi)\frac{dA_4}{dt}&=&3\alpha_t A_t+\alpha_h A_3
+5\alpha_4A_4+2\alpha_5A_5
+3\alpha_2M_2+\alpha_YM_Y+\frac{19}{6}\alpha_XM_X
+\frac{5}{6}\alpha_ZM_Z\theta_I  , \\
(2\pi)\frac{dA_5}{dt}&=&\alpha_h A_3
+2\alpha_4A_4+5\alpha_5A_5
+3\alpha_2M_2+\alpha_YM_Y+\frac{19}{6}\alpha_XM_X 
+\frac{5}{6}\alpha_ZM_Z\theta_I  ,\\
(2\pi)\frac{dA_k}{dt}&=&2\alpha_h A_3
+11\alpha_kA_k
+\frac{16}{3}\alpha_3M_3+\frac49\alpha_YM_Y+\frac{19}{6}\alpha_XM_X 
+\frac{5}{6}\alpha_ZM_Z\theta_I .
}
RGEs for scalar squared masses are given by
\eqn{
(2\pi)\frac{dm^2_{Q_a}}{dt}&=&\alpha_t M^2_t\delta_{a,3}
-\frac{16}{3}\alpha_3M^2_3-3\alpha_2M^2_2
-\frac19\alpha_YM^2_Y-\frac16\alpha_XM^2_X
-\frac{5}{18}\alpha_ZM^2_Z\theta_I  , \\
(2\pi)\frac{dm^2_{U^c_a}}{dt}&=&2\alpha_t M^2_t\delta_{a,3}
-\frac{16}{3}\alpha_3M^2_3
-\frac{16}{9}\alpha_YM^2_Y-\frac16\alpha_XM^2_X
-\frac{5}{18}\alpha_ZM^2_Z\theta_I  , \\
(2\pi)\frac{dm^2_{D^c_a}}{dt}&=&-\frac{16}{3}\alpha_3M^2_3
-\frac49\alpha_YM^2_Y-\frac23\alpha_XM^2_X
-\frac{10}{9}\alpha_ZM^2_Z\theta_I  , \\
(2\pi)\frac{dm^2_{L_a}}{dt}&=&-3\alpha_2M^2_2
-\alpha_YM^2_Y-\frac23\alpha_XM^2_X 
-\frac{10}{9}\alpha_ZM^2_Z\theta_I  , \\
(2\pi)\frac{dm^2_{E^c_a}}{dt}&=&
-4\alpha_YM^2_Y-\frac16\alpha_XM^2_X 
-\frac{5}{18}\alpha_ZM^2_Z\theta_I , \\
(2\pi)\frac{dm^2_{H^U_a}}{dt}&=&(3\alpha_tM^2_t+\alpha_h M^2_h
+2\alpha_4M^2_4)\delta_{a,3}+\alpha_5M^2_5(1-\delta_{a,3}) \no \\
&-&3\alpha_2M^2_2-\alpha_YM^2_Y-\frac23\alpha_XM^2_X 
-\frac{10}{9}\alpha_ZM^2_Z\theta_I  , \\
(2\pi)\frac{dm^2_{H^D_a}}{dt}&=&\alpha_h M^2_h\delta_{a,3}
+\alpha_4M^2_4(1-\delta_{a,3})+2\alpha_5M^2_5\delta_{a,3} \no \\
&-&3\alpha_2M^2_2-\alpha_YM^2_Y-\frac32\alpha_XM^2_X 
-\frac{5}{18}\alpha_ZM^2_Z\theta_I  , \\
(2\pi)\frac{dm^2_{S_a}}{dt}&=&(2\alpha_h M^2_h+9\alpha_kM^2_k)\delta_{a,3}
+(2\alpha_4M^2_4+2\alpha_5M^2_5)(1-\delta_{a,3}) \no \\
&-&\frac{25}{6}\alpha_XM^2_X
-\frac{5}{18}\alpha_ZM^2_Z\theta_I , \\
(2\pi)\frac{dm^2_{G_a}}{dt}&=&\alpha_kM^2_k-\frac{16}{3}\alpha_3M^2_3
-\frac49\alpha_YM^2_Y-\frac23\alpha_XM^2_X 
-\frac{10}{9}\alpha_ZM^2_Z\theta_I , \\
(2\pi)\frac{dm^2_{G^c_a}}{dt}&=&\alpha_kM^2_k-\frac{16}{3}\alpha_3M^2_3
-\frac49\alpha_YM^2_Y-\frac32\alpha_XM^2_X 
-\frac{5}{18}\alpha_ZM^2_Z\theta_I,
}
where
\eqn{
M^2_t&=&A^2_t+m^2_{Q_3}+m^2_{U^c_3}+m^2_{H^U_3} ,\quad
M^2_h=A^2_3+m^2_{S_3}+m^2_{H^U_3}+m^2_{H^D_3} ,\no \\
M^2_4&=&A^2_4+m^2_{S_1}+m^2_{H^U_3}+m^2_{H^D_1} ,\quad
M^2_5=A^2_5+m^2_{S_1}+m^2_{H^U_1}+m^2_{H^D_3} ,\no \\
M^2_k&=&A^2_k+m^2_{S_3}+m^2_{G_1}+m^2_{G^c_1}.  
}
Note that the relations
\eqn{
&&m^2_{S_1}=m^2_{S_2}=m^2_S,\quad
m^2_{H^U_1}=m^2_{H^U_2}=m^2_{H^U},\quad
m^2_{H^D_1}=m^2_{H^D_2}=m^2_{H^D},\quad
m^2_{Q_1}=m^2_{Q_2}=m^2_Q, \no \\
&&m^2_{L_1}=m^2_{L_2}=m^2_L,\quad 
m^2_{G_1}=m^2_{G_2}=m^2_{G_3}=m^2_G,\quad
m^2_{G^c_1}=m^2_{G^c_2}=m^2_{G^c_3}=m^2_{G^c},
}
are held. At $\mu=M_I$, we add $U(1)_Z$ D-term corrections as follows
\cite{dterm}
\eqn{
&&m^2_X(M_I-0)=m^2_X(M_I+0)+\Delta m^2_X \\
&&\Delta m^2_Q=\Delta m^2_{U^c}=\Delta m^2_{E^c}=\Delta m^2_{H^D}=\Delta m^2_{G^c}
=\Delta m^2_S=\frac{5}{18}m^2_{DT} ,\no \\
&&\Delta m^2_{D^c}=\Delta m^2_L=\Delta m^2_{H^U}=\Delta m^2_G=-\frac{5}{9}m^2_{DT} , \\
&&m^2_{DT}=1\mbox{TeV}^2>0.
}
We solve these RGEs using following boundary conditions.
At SUSY breaking scale ($\mu=M_S=1\mbox{TeV}$), we put by hand as
follows
\eqn{
&&\lambda_3=0.37,\quad \lambda_4=0.4,\quad \lambda_5=0.55 ,\quad
Y_t=Y^U_3=1.0 ,\quad k=0.5 ,\quad
M_3=1000\mbox{GeV}, \quad M_Y=200\mbox{GeV}, \no \\
&&m^2_{Q_3}=3.00 ,\quad m^2_{U^c_3}=1.00 ,\quad
m^2_{H^U}=m^2_{H^D}=m^2_S=2.00 ,\quad m^2_G=5.50 ,\quad
m^2_{G^c}=7.00\quad (\mbox{TeV}^2), \no \\
&&m^2_{H^U_3,H^D_3,S_3}\to \mbox{Eq.}(67)(68)(69).
}
At reduced Planck scale
($\mu=M_P=2.4\times 10^{18}\mbox{GeV}$),we put by hand as follows 
\eqn{
&&\alpha_2=\alpha_3=0.125 ,\quad \alpha_X=\alpha_Z=\alpha_Y=0.209,\quad
M_2=M_3,\quad M_X=M_Y=M_Z,\quad A_{t,3,4,5,k}=0, \no \\
&&m^2_{L_a}=m^2_{E^c_a}=m^2_{D^c_a}=m^2_{U^c_i}=m^2_Q=0 .
}
Note that gauge coupling constants do not satisfy the conventional unification as
\eqn{
\alpha_Y=\frac35\alpha_{2,3} .
}
The renormalization factors of first and second generation
Yukawa coupling constants $Y^{u,d,e}$ and
single G-Higgs coupling constants defined by 
\eqn{
W_G=Y^{QQ}Q_3Q_3(G_1+G_2+G_3)+Y^{UE}U^c_3E^c_3(G_1+G_2+G_3),
}
are given by
\eqn{
\sqrt{\frac{\alpha_A(M_S)}{\alpha_A(M_P)}},\quad \alpha_A=\frac{|Y^A|^2}{4\pi},\quad
A=u,d,e,QQ,UE ,
}
which are calculated by RGEs as follows
\eqn{
(2\pi)\frac{1}{\alpha_u}\frac{d\alpha_u}{dt}
&=&3\alpha_t+\alpha_h+2\alpha_4-\frac{16}{3}\alpha_3-3\alpha_2-\frac{13}{9}\alpha_Y
-\frac12\alpha_X-\frac{5}{6}\alpha_Z\theta_I  , \\
(2\pi)\frac{1}{\alpha_d}\frac{d\alpha_d}{dt}
&=&\alpha_h+2\alpha_5-\frac{16}{3}\alpha_3-3\alpha_2-\frac{7}{9}\alpha_Y
-\frac76\alpha_X-\frac{5}{6}\alpha_Z\theta_I , \\
(2\pi)\frac{1}{\alpha_e}\frac{d\alpha_e}{dt}
&=&\alpha_h+2\alpha_5-3\alpha_2-3\alpha_Y
-\frac76\alpha_X-\frac{5}{6}\alpha_Z\theta_I , \\
(2\pi)\frac{1}{\alpha_{UE}}\frac{d\alpha_{UE}}{dt}
&=&2\alpha_t+\alpha_k-\frac{16}{3}\alpha_3
-\frac{28}{9}\alpha_Y-\frac12\alpha_X-\frac56\alpha_Z\theta_I ,\\
(2\pi)\frac{1}{\alpha_{QQ}}\frac{d\alpha_{QQ}}{dt}
&=&2\alpha_t+\alpha_k-8\alpha_3-3\alpha_2
-\frac13\alpha_Y-\frac12\alpha_X-\frac56\alpha_Z\theta_I ,
}
where the contributions from $\Phi_a,\Phi^c_3,D_i$ are neglected.
The results are given in Table 6.

\begin{table}[htbp]
\begin{center}
\begin{tabular}{|c|c|c|c|c|c|}
\hline
parameter  &$\mu=M_S(M_I)$&$\mu=M_P$&parameter     &$\mu=M_S$     &$\mu=M_P$ \\ \hline
$\alpha_Y$ &0.010442      &0.209    &$m^2_{Q_3}$   &3.00          & 0.9912   \\ \hline
$\alpha_2$ &0.032482      &0.125    &$m^2_{U^c_3}$ &1.00          & 3.1451   \\ \hline
$\alpha_3$ &0.089430      &0.125    &$m^2_{H^U_3}$ &$-0.1723$     &16.4770   \\ \hline
$\alpha_X$ &0.010552      &0.209    &$m^2_{H^U}$   &2.00          & 2.4671   \\ \hline
$\alpha_Z$ &(0.015162)    &0.209    &$m^2_{H^D_3}$ &2.6811        & 6.2912   \\ \hline
$\alpha_t$ &0.079577      &0.006455 &$m^2_{H^D}$   &2.00          & 1.6459  \\ \hline
$\alpha_h$ &0.010894      &0.016086 &$m^2_{S_3}$   &$-2.2105$     &10.4216   \\ \hline
$\alpha_4$ &0.012732      &0.011761 &$m^2_S$       &2.00          & 6.4914   \\ \hline
$\alpha_5$ &0.024072      &0.011309 &$m^2_G$       &5.50          & 1.6852   \\ \hline
$\alpha_k$ &0.019894      &0.001014 &$m^2_{G^c}$   &7.00          & 2.2501   \\ \hline
$M_Y$      &0.2           &3.68582  &$m^2_Q$       &5.9677        &0.0 \\ \hline
$M_2$      &0.49889       &1.68366  &$m^2_{U^c_i}$ &5.7726        &0.0 \\ \hline
$M_3$      &1.0           &1.68366  &$m^2_{D^c_a}$ &4.8538        &0.0 \\ \hline
$M_X$      &0.20228       &3.68582  &$m^2_{L_a}$   &1.0559        &0.0 \\ \hline
$M_Z$      &(0.28352)     &3.68582  &$m^2_{E^c_a}$ &1.7796        &0.0 \\ \hline 
$A_t$      &$-1.78127$    &0.0      &$\alpha_u$    &0.259802      &0.01 \\ \hline
$A_3$      &1.51978       &0.0      &$\alpha_d$    &0.523940      &0.01 \\ \hline
$A_4$      &0.40541       &0.0      &$\alpha_e$    &0.037691      &0.01 \\ \hline
$A_5$      &$-0.93253$    &0.0      &$\alpha_{UE}$ &0.238230      &0.01 \\ \hline
$A_k$      &$-3.05177$    &0.0      &$\alpha_{QQ}$ &1.638921      &0.01 \\ \hline
\end{tabular}
\end{center}
\caption{Each boundary values of the solutions of RGEs. The dimensionful parameters
are expressed in TeV units. The experimental values of gauge coupling constants 
give $\alpha_Y(M_S)=0.010445, \alpha_2(M_S)=0.032484,
\alpha_3(M_S)=0.089514$ which are calculated based on SM RGEs.
The values of $\alpha_Z,M_Z$ at low energy side are given by the values at $\mu=M_I$
in brackets.}
\end{table}

\section{The multiplication rules of $S_4$}

The representations of $S_4$ are $1,1',2,3,3'$ \cite{review}.
Their products are expanded as follows.
\begin{eqnarray}
\3tvec{x_1}{x_2}{x_3}_3\times\3tvec{y_1}{y_2}{y_3}_3
&=&\L(x_1y_1+x_2y_2+x_3y_3\R)_1
+\2tvec{\sqrt{3}(x_2y_2-x_3y_3)}{(x_2y_2+x_3y_3-2x_1y_1)}_2
+\3tvec{x_2y_3-x_3y_2}{x_3y_1-x_1y_3}{x_1y_2-x_2y_1}_{3'} \no \\
&+&\3tvec{x_2y_3+x_3y_2}{x_3y_1+x_1y_3}{x_1y_2+x_2y_1}_3
=\3tvec{x_1}{x_2}{x_3}_{3'}\times\3tvec{y_1}{y_2}{y_3}_{3'} , \\
\3tvec{x_1}{x_2}{x_3}_3\times\3tvec{y_1}{y_2}{y_3}_{3'}
&=&\L(x_1y_1+x_2y_2+x_3y_3\R)_{1'}
+\2tvec{(x_2y_2+x_3y_3-2x_1y_1)}{-\sqrt{3}(x_2y_2-x_3y_3)}_2 \no \\
&+&\3tvec{x_2y_3-x_3y_2}{x_3y_1-x_1y_3}{x_1y_2-x_2y_1}_3
+\3tvec{x_2y_3+x_3y_2}{x_3y_1+x_1y_3}{x_1y_2+x_2y_1}_{3'} , \\
\2tvec{x_1}{x_2}_2\times\3tvec{y_1}{y_2}{y_3}_{3(3')}
&=&\3tvec{2x_2y_1}{-\sqrt{3}x_1y_2-x_2y_2}
{\sqrt{3}x_1y_3-x_2y_3}_{3(3')}
+\3tvec{2x_1y_1}{-x_1y_2+\sqrt{3}x_2y_2}
{-x_1y_3-\sqrt{3}x_2y_3}_{3'(3)} , \\
\3tvec{x_1}{x_2}{x_3}_{3(3')}\times (y)_{1'}&=&
\3tvec{x_1y}{x_2y}{x_3y}_{3'(3)} , \\
\2tvec{x_1}{x_2}_2\times\2tvec{y_1}{y_2}_2
&=&(x_1y_1+x_2y_2)_1+(x_1y_2-x_2y_1)_{1'}
+\2tvec{x_1y_2+x_2y_1}{x_1y_1-x_2y_2}_2 , \\
\2tvec{x_1}{x_2}_2\times (y)_{1'}&=&\2tvec{-x_2y}{x_1y}_2 , \\
(x)_{1'}\times(y)_{1'}&=&(xy)_1.
\end{eqnarray}

\section{Mass bounds of new particles}

\begin{table}[htbp]
\begin{center}
\begin{tabular}{|c|c|c|c|c|c|}
\hline
particle            &mass   &exp                  &particle      &mass &exp    \\ \hline
$H^0(\mbox{lightest even})$ &125.7  &$125-126$\cite{higgs}&$\chi^\pm_3$  &1486 &$>295$ \cite{atlas} \\ \hline
$T_+$               &1882   &$>560$\cite{atlas}   &$\chi^\pm_w$  &493  &$>295$ \cite{atlas} \\ \hline
$T_-$               &1178   &$>560$\cite{atlas}   &$\chi^0_1$    &199  &$>46$  \cite{PDG2012} \\ \hline
$G_+$               &3908   &                     &$\chi^0_2$    &493  &$>62.4$\cite{PDG2012}\\ \hline
$G_-$               &1737   &$(>683)$\cite{atlas} &$\chi^0_3$    &1481 &$>99.9$\cite{PDG2012}\\ \hline
$Q_{1,2}$       &$\geq 2532$&$>1380$\cite{atlas}  &$\chi^0_4$    &1487 &$>116$ \cite{PDG2012} \\ \hline
$U^c_{1,2}$     &$\geq 2493$&$>1380$\cite{atlas}  &$\chi^0_5$    &2004 &       \\ \hline
$D^c_{1,2,3}$   &$\geq 2395$&$>1380$\cite{atlas}  &$\chi^0_6$    &2208 &       \\ \hline
$L_{1,2,3}$     &$\geq 1393$&$>195$\cite{atlas}   &$\chi^\pm_i$  &119  &$100-140$\cite{lep2}\cite{atlas}\\ \hline
$E^c_{1,2,3}$   &$\geq 1490$&$>195$\cite{atlas}   &$\chi^0_{i,1}$&36.6 &       \\ \hline
$H^U_{1,2}(even,odd,\pm)$&1056   &$>93.4$\cite{PDG2012}&$\chi^0_{i,2}$&120  &$>116$ \cite{PDG2012}  \\ \hline
$H^D_{1,2}(even,odd,\pm)$&821    &$>93.4$\cite{PDG2012}&$\chi^0_{i,3}$&157  &$>116$ \cite{PDG2012}  \\ \hline
$S_{1,2}(even,odd)$      &2052   &                     &$g$           &2000 &       \\ \hline
$H_3(even,odd,\pm)$      &2279   &$>93.4$\cite{PDG2012}&$\lambda^g_3$ &1000 &$>1000$\cite{atlas}  \\ \hline 
$S_3(even)$            &2102   &                     &$Z'$          &2102 &$>1520$\cite{zprime2011}\\ \hline
\end{tabular}
\end{center}
\caption{Mass values of new particles calculated based on our assumption and
corresponding experimental constraints in GeV units. The capital letters means bosons and the Greek
characters and the small letter mean fermions. The equations
which are used to calculate mass values, are Eq.(10),(71),(74),(75),(76),(77),
(83),(85),(95),(96),(97),(98). Each equalities in "mass" column correspond to imposing
the boundary conditions as $m^2_X=0 (X=Q,U^c_i,D^c_a,L_a,E^c_a)$ at $\mu=M_P$.
We adopt the mass bound for stable stop as one for lighter G-Higgs ($G_-$) in bracket,
under the assumption that $G_-$ is lighter than $g$ and $G_+$.
We adopt the mass bound for CP-odd Higgs boson in supersymmetric model
as ones for extra Higgs bosons ($H^U_{1,2},H^D_{1,2},H_3$).}
\end{table}

The mass bound of the lightest chargino ($\chi^\pm_1$) is given by 3-lepton emission 
through EW direct process $\chi^\pm_1\chi^0_2\to W^\pm Z\chi^0_1\chi^0_1$.
This neutralino $\chi^0_1$ corresponds to $\chi^0_{i1,}$ in our model.
Under the assumption that  slepton decouples and LSP ($\chi^0_1$) is massless,
excluded region of chargino mass is given by 
$140<M(\chi^\pm_1)<295\mbox{GeV}$ \cite{atlas},
or $M(\chi^\pm_1)<330\mbox{GeV}$ \cite{cms}.
These constraints are not imposed on the chargino
in the mass range
\eqn{
M(\chi^\pm_1)=M(\chi^0_2)<M(\chi^0_1)+m_Z.
}
In this case $Z$ and following two lepton emissions are suppressed.
Taking account of LEP bound $M(\chi^\pm_1)>100\mbox{GeV}$ \cite{lep2},
we consider the allowed region given by
\eqn{
100<M(\chi^\pm_1)<140 \quad (\mbox{GeV}).
}



\end{document}